\documentclass[12pt,preprint,showpacs,tightenlines,preprintnumbers,nofootinbib]{revtex4}
\usepackage{epsf,graphicx}
\usepackage{amsmath}

\def\beq{\begin{equation}}
\def\eeq{\end{equation}}
\def\bea{\begin{eqnarray}}
\def\eea{\end{eqnarray}}
\def\beqa{\begin{equation}\begin{array}{l}}
\def\eeqa{\end{array}\end{equation}}
\def\eqlab#1{\label{eq:#1}}
\def\figlab#1{\label{fig:#1}}

\def\barr{\left(\begin{array}{c}}
\def\earr{\end{array}\right)}
\def\bmat{\left(\begin{array}{cc}}
\def\emat{\end{array}\right)}
\def\eref#1{(\ref{eq:#1})}
\def\Eqref#1{Eq.~(\ref{eq:#1})}

\def\Figref#1{Fig.~\ref{fig:#1}}

\def\sla#1{#1  \!\!\!\!\slash}
\def\slaa{a \hspace{-2.25mm} \slash}
\def\slap{p \hspace{-2mm} \slash}

\def\half{\mbox{\small{$\frac{1}{2}$}}}

\def\thalf{\mbox{\small{$\frac{3}{2}$}}}
\def\quarter{\mbox{\small{$\frac{1}{4}$}}}
\def\third{\mbox{\small{$\frac{1}{3}$}}}

\def\al{\alpha}
\def\be{\beta}
\def\ga{\gamma} \def\Ga{{\it\Gamma}}
\def\de{\delta} \def\De{\Delta}
\def\veps{\varepsilon}  \def\eps{\epsilon}

\def\la{\lambda} \def\La{{\Lambda}}

\def\si{\sigma} \def\Si{{\it\Sigma}}
  \def\Th{\Theta}

\def\pa{\partial}

\def\pa{\partial}

\def\nn{\nonumber}

\def\Tau{{\mathcal T}}
\def\lag{{\mathcal L}}

\def\MM{{\mathcal M}}

\def\mathscr{\mathcal}
\def\N{N}

\def\3d{3-D}

\def\ol#1{\overline{#1}}

\def\ceft{$\chi$EFT}
\def\EM{e.m.}

\begin{document}

\title{Chiral Effective-Field Theory in the
$\Delta$(1232) Region:\\ 
Pion Electroproduction on the Nucleon}

\author{Vladimir Pascalutsa}
\email{vlad@jlab.org}
\author{Marc Vanderhaeghen}
\email{marcvdh@jlab.org}

\affiliation{Physics Department, The College of William \& Mary, Williamsburg, VA
23187, USA\\
Theory Center, Jefferson Lab, 12000 Jefferson Ave, Newport News, 
VA 23606, USA}

\date{\today}

\begin{abstract}
We develop an extension of chiral perturbation theory to the
$\Delta$(1232)-resonance energy region and apply it to investigate
the pion electroproduction 
off the nucleon ($e^-\, N \rightarrow e^-\, N\,\pi$).
We present a complete calculation of this process,
in the $\Delta$-resonance region, up to next-to-leading
order in the  {\it $\de$-expansion}. 
At this order, the only free parameters are the three
low-energy constants  corresponding to the magnetic (M1), electric (E2),
and Coulomb (C2) $\gamma N \to \Delta $ transition strength.
After fitting these parameters to a few well-known data, our
calculation provides a prediction for observables and multipole
amplitudes of pion electroproduction. These results compare
favorably with the phenomenological multipole solutions
and recent experimental results from MIT-Bates and MAMI. 
Our prediction for the pion-mass dependence of the
$\ga N\De$ form factors offers an explanation for the discrepancy
between the recent lattice-QCD results and the experimental 
value for the ``C2/M1 ratio" at low $Q^2$.   
\end{abstract}


\maketitle
\thispagestyle{empty}

\section{Introduction}

The first excited state of the nucleon --- the $\Delta(1232)$ resonance ---  
dominates pion-production phenomena and plays an important role in 
our understanding of the low-energy nucleon structure. 
High-precision measurements of the nucleon-to-$\De$
transition by means of electromagnetic probes became possible with the
advent of the new generation of electron scattering facilities, such as 
BATES, MAMI, and JLab,  many measurements being completed
in recent years~\cite{Mainz97,LEGS97,Bates01,Jlab}.

The {\it electromagnetic}  nucleon-to-$\De$ (or, in short $\ga N \De$)  
transition is predominantly
of the magnetic dipole ($M1$) type. In a simple quark-model picture, 
this $M1$ transition 
is described by a spin flip of a quark in the $s$-wave state. 
Any $d$-wave admixture
in the nucleon {\it or} the $\Delta$ wave-functions allows for the electric ($E2$) and Coulomb ($C2$)
quadrupole transitions. Therefore by measuring these one is able to assess the presence
of the $d$-wave components and hence quantify to which extent the nucleon or the $\De$ wave-function
deviates from the spherical shape, {\it i.e.}, to which extent they are 
``deformed''~\cite{deformation}. 
The  $d$-wave component of $\De$'s wave-function can be separately
assessed by measuring its electric quadrupole moment. However, 
this would be extremely difficult because of the tiny lifetime of the $\De$.
The $\ga N\Delta$ transition, on the other hand,  was
accurately measured in the pion photo- and electro-production reactions 
in the $\De$-resonance energy region. The $E2$ and $C2$ transitions 
 were found to be relatively small at moderate  momentum-transfers ($Q^2$), 
the ratios $R_{EM}=E2/M1$ and 
$R_{SM}=C2/M1 $ are at the level of a few percent.  

Traditionally, the resonance parameters are extracted 
by using {\it unitary isobar models}~\cite{ELAs,SpainModel,Drechsel:2000um,KVItheory,GiessenModel,Drechsel:1998hk,Inna}, which
in essence are unitarized tree-level calculations based on
phenomenological Lagrangians. However, 
at low  $Q^2$ the $\gamma N \Delta$-transition 
shows great sensitivity to the ``pion 
cloud'', which until recently could only be
comprehensively studied within 
{\it dynamical models}~\cite{Sato2, GrS96, DMT,Fuda03,PaT00,Caia04}.
(Unlike the isobar models, dynamical models include quantum effects
due to pion loops.)

With the advent of the chiral effective field theory (\ceft) 
of QCD~\cite{Weinberg:1978kz,Gasser:1983yg}
and its extensions to the $\De$-resonance region~\cite{isospurion,PP03}, 
it has become possible to study the nucleon and $\Delta$-resonance properties 
in a profoundly different way. Recently,
we have been able to perform  first \ceft\  studies of the 
$\gamma N \Delta$-transition 
in pion electroproduction~\cite{Pascalutsa:2005ts} and 
of the $\De$-resonance magnetic moment in the radiative
pion photoproduction~\cite{PV05}. The advantages over the
previous dynamical approaches are apparent: \ceft\ is
a low-energy effective field theory of QCD and as such 
it provides a firm theoretical foundation, with all
the relevant symmetries and scales of QCD built in consistently.
Moreover, we find that already at next-to-leading order 
(NLO) in the ``$\de$-expansion'' of Ref.~\cite{PP03},  the observables
for pion electroproduction in the $\De$-resonance region 
are described remarkably well. The \ceft, therefore, 
provides a theoretically
consistent and phenomenologically viable  framework, which, in particular, 
will allow for a model-independent extraction of the resonance parameters.

Tremendous progress has recently been achieved as well in the 
lattice QCD simulations of the $\ga N\De$ transition. The present
state-of-the-art results~\cite{Ale05} are ``quenched'' and are obtained
for pion masses above 300 MeV.
These results can only  be confronted with experiment after
an extrapolation  down to the physical pion mass of $140$ MeV.
A {\it linear} in the 
quark mass ($m_q \sim m_\pi^2$) extrapolation was used in Ref.~\cite{Ale05}.
The thus obtained $R_{SM}$ ratio, at low $Q^2$, was found to be  
in major disagreement with experiment. 
The apparent caveat of this result is that the extrapolation 
in the quark mass needs not to be linear. The
non-analytic dependencies, such as $\sqrt{m_q}$
and $\ln m_q$, are known to be important as one approaches the
small physical value of $m_q$.  
These  non-analytic terms can also be obtained from \ceft,
and, as we have demonstrated~\cite{Pascalutsa:2005ts},
the  $R_{EM}$ and $R_{SM}$ ratios 
do exhibit a pronounced non-analytic quark-mass dependence, such that
the lattice results~\cite{Ale05} can be reconciled with experiment.
Here we shall refine our calculation by consistently including
the quark-mass dependence of the nucleon and $\De$-isobar masses
in the same \ceft\ framework~\cite{PV05self}.

 A brief account of this work has recently been
published~\cite{Pascalutsa:2005ts}. In
 this paper we present an extensive description of the work,
improve on our theoretical error analysis, and
present some new results that could 
not be included in the brief publication.
This paper is organized as follows.

In Sec.~\ref{sec2} we recall the relevant chiral Lagrangians,
while Sec.~\ref{sec3} explains the power counting in the $\de$-expansion
scheme based on which 
the leading- and next-to-leading-order contributions are selected. 
The chiral-loop contributions 
to the $\gamma N \Delta$-transition
 form factors are evaluated in Sec.~\ref{sec5} by using two
different techniques. 
In Sec.~\ref{sec4} we discuss the theoretical uncertainty of
our calculation due to the neglect of higher-order effects.  
In Sec.~\ref{sec6} we present the results for 
pion photo- and electroproduction observables, multipoles,
and the extracted $\ga N\De$ form factors. In that section we
also discuss the \ceft\ predictions for
the $m_\pi$-dependence of the $\ga N\De$ transition and compare them
with available lattice results. 
Sec.~\ref{sec7} lists the main points and conclusions of the paper.
The two Appendices contain technical details concerning the
Feynman-parameter and dispersion integrals, respectively.

\section{The effective chiral Lagrangian}
\label{sec2}

In this section we define the relevant effective Lagrangian
of low-energy QCD, where we restrict ourselves to the two flavor, 
isospin symmetric case ($m_u = m_d\equiv m_q$). The guiding principle
is the symmetry under the SU(2)$_L\times$SU(2)$_R$ chiral rotations,
see \cite{Weinberg:1978kz,Gasser:1983yg,GSS89}.
The lowest-order Lagrangian of the Goldstone-boson (pion) isovector field
$\pi^a$ ($a=\ol{1,3}$)
\beq
\lag_\pi^{(2)} = \quarter f_\pi^2\, \mbox{Tr}[ \,\pa_\mu  U
\pa^\mu U^\dagger +  ( U + U^\dagger )\,\chi\, ]
= \half \pa_\mu \pi^a \pa^\mu \pi^a - \half m_\pi^2 [\, \pi^2 + O(\pi^4)\,] , 
\eeq  
where 
\beq
U(x) = e^{i\pi^a(x)\,\tau^a/f_\pi} \equiv u^2(x), 
\,\,\,\, \mbox{with} \,\,\,
\pi^a \tau^a = 
\bmat \pi^0  &\sqrt{2}\,\pi^+ \\
\sqrt{2}\,\pi^- &  - \pi^0  \emat,
\eeq
and the linear in $U$, explicit 
symmetry-breaking term is proportional to the quark (or, pion)
masses:  $\chi = 2B m_q = m_\pi^2 + O(m_\pi^4)$.
Furthermore, at this lowest order, $f_\pi \simeq 92.4$ MeV is the pion decay
constant and $B$ is related to the scalar quark condensate as $B=
-\left< \bar q q\right>/f_\pi^2 $. 
In the notation $\lag^{(i)} $, the superscript stands for the number of
derivatives of Goldstone-boson fields and insertions of their mass.

The chirally symmetric Lagrangian for the nucleon isodoublet 
field, $N=(p,n)^T$, can conveniently
be written by using the SU(2) vector and axial-vector currents,
\begin{subequations}
\eqlab{currents}
\bea
\half\, \tau^a v_\mu^a(x) & \equiv &  \frac{1}{2i} \left(u \,
\pa_\mu u^\dagger+u^\dagger \pa_\mu  u \right)
= \frac{1}{4 f_\pi^2}  \, \veps^{abc} \, \tau^a \pi^b \, \pa_\mu \pi^c  
+ O(\pi^4), \\
\half \,\tau^a a^{\,a}_\mu(x) & \equiv &  \frac{1}{2i} \left(u^\dagger \,
\pa_\mu  u- u \,\pa_\mu  u^\dagger \right) 
= \frac{1}{2 f_\pi} \tau^a \,\pa_\mu  \pi^a  + O( \pi^3), 
\eea
\end{subequations}
and the chiral covariant derivative acting on the nucleon field, 
\beq 
D_\mu N = \pa_\mu N  + \half i \tau^a v_\mu^a\, N .
\eeq  
In terms of these definitions the nucleon Lagrangian begins at
\bea
\eqlab{Nlagran}
\lag^{(1)}_N &=& \ol N( i \sla{D} -{M}_{N} +  \half g_A \,  
\tau^c \slaa^{\,c} \ga_5 ) N\,,
\eea
where $g_A\simeq 1.267$ is the nucleon axial-coupling constant.

In the presence of the electromagnetic field ($A_\mu$), the charge
of the pions is accounted for by making the ``minimal substitution'':
$\pa_\mu \pi^a  \to \pa_\mu \pi^a + e  \,\veps^{ab3}A_\mu\pi^b $,
in the above expressions.  Similarly, the proton charge is included
by the minimal substitution in the chiral derivative as: 
$D_\mu N \to D_\mu N + ie \half (1+\tau^3) A_\mu N$.

The Lagrangian for the spin-3/2 isospin-3/2 $\De$-isobar can be written 
in terms of the Rarita-Schwinger (vector-spinor) {\it isoquartet} field\footnote{Frequently chiral Lagrangians for the $\De$-isobar are
written in terms of the isovector-isodoublet field~\cite{isospurion}:
$\De_\mu^{a}  =  T^a \De_\mu$, with $T$ isospin-1/2-to-3/2 
transition matrices. It is an  alternative to the isoquartet
representation adopted in this work. The two representations, however,
are equivalent at the level of observables.},
$\De_\mu = (\De_\mu^{++},\De_\mu^{+}, \De_\mu^{0}, \De_\mu^{-})^T $, 
by using the vector and axial-vector currents \Eqref{currents} in the
isospin-3/2 representation of SU(2). The corresponding generators $\Tau^a$
have the following matrix representation:
\begin{subequations}
\bea
\Tau^1 & = & \frac{2}{3} \left(\begin{array}{cccc}
0 & \sqrt{3}/2  & 0 & 0 \\
\sqrt{3}/2 & 0 & 1 & 0 \\
0 & 1 & 0 & \sqrt{3}/2\\
0 & 0 & \sqrt{3}/2 & 0 
\emat, \\
\Tau^2 & = & \frac{2i}{3} \left(\begin{array}{cccc}
0 & -\sqrt{3}/2  & 0 & 0 \\
\sqrt{3}/2 & 0 &-1 & 0 \\
0 & 1 & 0 & -\sqrt{3}/2\\
0 & 0 & \sqrt{3}/2 & 0 
\emat,\\
\Tau^3 & = & \mbox{diag} (1, \third, -\third, -1),
\eea
\end{subequations}
and satisfy $\Tau^a \Tau^a = 5/3$. 
The chiral derivative is then given by:
\beq 
D_\mu \De_\nu = \pa_\mu \De_\nu  + 
 i \,v_\mu^a\,\Tau^a  \De_\nu ,
\eeq
and the  first-order chiral Lagrangian can be written as:
\beq
\eqlab{deltaLag}
\lag^{(1)}_\De = \ol\De_\mu \left(i\ga^{\mu\nu\rho}\,D_\rho - 
M_\De\,\ga^{\mu\nu} \right) \De_\nu - 
 \half H_A  \,\ol\De_\mu 
\, \slaa^{\,c}\,\Tau^c \ga_5\, \De^\mu,
\eeq
where  $M_\De\simeq 1.232 $ GeV is the mass of the $\De$-isobar,
$H_A$ is the axial coupling constant given, 
in the large-$N_C$ limit~\cite{Dashen:1993jt},
by $H_A=(9/5) g_A$ and this value is known to be consistent
with the empirical information, see, e.g.,~\cite{Butler:1992pn}. 
The totally-antisymmetric products
of $\ga$-matrices are defined as: $\ga^{\mu\nu}=\half[\ga^\mu,\ga^\nu]$,
$\ga^{\mu\nu\al}= i\veps^{\mu\nu\al\be}\ga_\be\ga_5$. The electric
charge of the $\De$ is accounted for by the following
minimal substitution: $D_\mu \to D_\mu + ie \half(1+3\Tau^3) A_\mu$.

The free Rarita-Schwinger field obeys the following field equation:
\beq
\eqlab{fieldeq}
i\ga^{\mu\nu\rho}\,\pa_\rho \De_\nu = M_\De  \ga^{\mu\nu} \De_\nu,
\eeq
which, in particular, yields  the constraints:
$ \ga^\mu \De_\mu = 0 =   \pa^\mu \De_\mu $. The constraints reduce the
number of  spin degrees of freedom of the vector-spinor
field to the physical number appropriate for a massive particle with
spin 3/2. These constraints arise essentially due to the invariance
of the free-field kinetic term [l.h.s.\ of \Eqref{fieldeq}], 
upon the gauge transformation
of the spin-3/2 field:
\beq
\eqlab{gsym}
\De_\mu(x) \to \De_\mu(x) +\pa_\mu \eps(x),
\eeq 
where $\eps$ is a spinor. In order for interactions to
support the number of the free-field constraints, they must be 
symmetric with respect to the same (or, at least,  similar) 
transformation~\cite{Pas98}. Now, the chiral interactions
in \Eqref{deltaLag} obviously do not have this symmetry. Therefore,
we redefine them by using the free field equation, or, equivalently,
redefine the spin 3/2 field, such that they become symmetric~\cite{Pas01}.
Then, e.g., the lowest-order axial coupling becomes
\beq
\lag^{(1)}_{\De\De\pi} = -\frac{H_A}{2M_\De f_\pi}  \,
\veps^{\mu\nu\rho\si} \,\ol\De_\mu 
\, \Tau^a \,(\pa_\rho \De_\nu )\,\pa_\si \pi^{a} + O(\pi^3).
\eeq
The difference with the original interaction term from \Eqref{deltaLag}
is of higher order (in the pion derivatives and mass). 

We write the $N\De$-transition Lagrangian right away in such an expanded
form which exhibits the spin-3/2 gauge symmetry:
\begin{subequations}
\eqlab{lagran}
\bea
\lag^{(1)}_{N\De} &=&  \frac{i h_A}{2 f_\pi M_\De}
\ol N\, T^a \,\ga^{\mu\nu\la}\, (\pa_\mu \De_\nu)\, \pa_\la \pi^a 
+ \mbox{H.c.}, \\
\lag^{(2)}_{N\De} &=&   \frac{3 i e g_M}{2M_N (M_N + M_\Delta)}\,\ol N\, T^3
\,\pa_{\mu}\De_\nu \, \tilde F^{\mu\nu}  + \mbox{H.c.},\\
\lag^{(3)}_{N\De} &=&  \frac{-3 e}{2M_N (M_N + M_\Delta)} \ol N \, T^3
\ga_5 \left[ g_E (\pa_{\mu}\De_\nu) 
+  \frac{i g_C}{M_\De} \ga^\al  
(\pa_{\al}\De_\nu-\pa_\nu\De_\al) \,\pa_\mu\right] F^{\mu\nu}+ \mbox{H.c.},
\;\;\;\;\; 
\eea
\end{subequations}
where $F^{\mu\nu}$ and $\tilde F^{\mu\nu}$
are the electromagnetic field strength and its dual,
$T^a$ are the isospin-1/2-to-3/2 transition $(2\times 4$) matrices:
\begin{subequations}
\bea
T^1 & = & \frac{1}{\sqrt{6}} \left(\begin{array}{cccc}
-\sqrt{3}   & 0 & 1 & 0 \\
0 & -1 & 0 & \sqrt{3}  
\emat, \\
T^2 & = & \frac{-i}{\sqrt{6}} \left(\begin{array}{cccc}
\sqrt{3}  & 0 & 1 & 0 \\
 0 & 1 & 0 & \sqrt{3}
\emat,\\
T^3 & = & \sqrt{\frac{2}{3} } \left(\begin{array}{cccc}
 0 & 1 & 0 & 0\\
 0 & 0 & 1 & 0  
\emat,
\eea
\end{subequations}
satisfying $T^a T^{b\dagger} = \de^{ab} -\third \tau^a \tau^b$.

Note that the electric and the Coulomb $\ga N\De$ couplings 
are of one order higher than
the magnetic one, because of the $\ga_5$ which involves 
the ``small components'' of the fermion fields and thus 
introduces an extra power of the 3-momentum.

For the momentum-space $\De$ propagator we  use
\beq
S^{\al\be} (p) = \frac{\slap+M_\De}{M_\De^2 - p^2}\left[ g^{\al\be} -\third 
\ga^\al \ga^\be + \frac{(1-\zeta)(\zeta \slap +M_\De)}{3(\zeta^2 p^2-M_\De^2)}
(\ga^\al p^\be - \ga^\be p^\al) 
+ \frac{2 (1-\zeta^2)\,p^\al p^\be}{3(\zeta^2 p^2-M_\De^2)} \right],
\eeq
where $\zeta$ is a gauge-fixing parameter,
see Refs.~\cite{PV05self,Pa98thesis}
for details. The analog of the Landau-gauge for spin-1 case here is, 
$\zeta = \infty$:
\beq
 S^{\al\be} (p)=
\frac{\slap+M_\De}{M_\De^2 - p^2} \, {\mathscr P}^{(3/2)\,\al\be}(p),
\eeq 
with 
\beq
{\mathscr P}^{(3/2)\,\al\be}(p) =\frac{2}{3} 
\left(g^{\al\be} - \frac{p^\al p^\be}{p^2} \right) +\frac{\slap}{3 p^2} 
\ga^{\al\be\mu}\, p_\mu ,
\eeq
the covariant 
spin-3/2 projection operator.
As long as the coupling of the spin-3/2 field are gauge-symmetric
with respect to the transformation \eref{gsym}, the results are independent
of the choice of the gauge-fixing parameter $\zeta$.

\section{Power counting and renormalization}
\label{sec3}

The inclusion of the 
$\De$-resonance introduces another light scale in the theory,
the resonance excitation energy: $\De\equiv M_\De-M_N\sim 0.3$ GeV.
This energy scale is still relatively light in comparison to
the chiral symmetry breaking scale
$\La_{\chi SB} \sim 1$ GeV. Therefore,
$\de = \De/\La_{\chi SB}$ can be treated as a small parameter. 
The question is, how to compare this parameter
with the small parameter of chiral perturbation theory ($\chi$PT),
$\eps = m_\pi /\La_{\chi SB}$. After all, our aim is to organize
an expansion in {\it a} small parameter and to estimate the size of various
contributions based on power-counting rules.  So some
specific relation between $\de$ and $\eps$ would be helpful.
 
In most of the literature (see, e.g.,~\cite{isospurion}) 
they are assumed to be of 
comparable size, $\de\approx \eps$. 
This, however, leads to a somewhat unsatisfactory
result: the $\De$-resonance 
contributions are always estimated to be of the
same size as the nucleon contributions. Thus, obviously,
the $\De$-contributions
are {\it overestimated} at lower energies and {\it underestimated}
at the resonance energies. To estimate the $\De$-resonance
contributions correctly, and depending on the energy region,
one needs to count $\de$ and $\eps$ {\it differently}.

A relation $\eps =\de^2$ was suggested and explored in Ref.~\cite{PP03},
and is adopted in this work.
The second power is indeed the closest integer power for
the relation of these parameters in the real world.
We should stress that this relation is used for power-counting purposes
only. It is not imposed in the actual evaluations of diagrams.
Each diagram is simply  characterized 
by an overall $\delta$-counting index $n$,
which tells us that its contribution begins at O($\de^n$). 

Because of the distinction of $m_\pi$ and $\De$ the counting of a given diagram depends 
on whether the characteristic momentum $p$ is  
in the low-energy region ($p\sim m_\pi$) or in the resonance
region ($p\sim \De$). 
In the low-energy region the index of a graph with $L$ loops, $N_\pi$ pion propagators, $N_N$
nucleon propagators, $N_{\De}$ $\Delta$-propagators, and $V_i$ vertices of
dimension $i$ is 
\beq
n= 
2 \,(\,\sum_i i V_i + 4 L  - N_N - 2 N_\pi ) 
-N_\De \equiv 2 n_{\chi{\mathrm PT}} - N_\De, 
\eeq
where $ n_{\chi{\mathrm PT}} $
is the index in $\chi$PT with no $\De$'s \cite{GSS89}.
In the resonance region, one distinguishes the one-$\De$-reducible (O$\De$R) graphs~\cite{PP03}, see e.g.,
graph (a) in \Figref{diagrams}. Such graphs contain $\De$ propagators
which go as $1/(p-\De)$, and hence for $p\sim \De$ they are large and 
all need to be included. This gives an incentive, {\it within
the power-counting scheme}, to 
resum $\De$ contributions.
Their resummation amounts to 
dressing the $\De$ propagators so that they behave as $1/(p-\De-\Si)$. The self-energy 
$\Si$ begins at order $p^3$ and thus
a dressed O$\De$R propagator counts as $1/\de^3$.
If the number of
such propagators in a graph is $N_{O\De R}$, the power-counting index of
this graph in the resonance region is given by 
\beq
n=n_{\chi{\mathrm PT}} - N_\De - 2N_{O\De R},
\eeq 
where $N_\De$ is the total number
of $\De$-propagators.

A word on the renormalization program, as it is an indivisible part
of power counting in a relativistic theory. Indeed, without some
kind of renormalization the loop graphs diverge as $\La^{\cal N} $,
where $\La$ is an ultraviolet cutoff, and ${\cal N}$ is a positive
power proportional to the power-counting index of the graph. 
Also, contributions of heavy scales,
such as baryon masses, may appear as $M^{\cal N}$. 
The renormalization
of the loop graphs can and should be performed so as to absorb
these large contributions into the available low-energy constants,
thus bringing the result in accordance
with power counting \cite{Gegelia:1999gf}.

To give an example, consider the one-$\pi N$-loop contribution to
the nucleon mass, \Figref{nucself}. For the $\pi NN$ vertex from
$\lag^{(1)}_N$ the power counting tells us that this contribution
begins at $O(m_\pi^3)$. An explicit calculation, however, will 
show (e.g., \cite{GSS89}) that the loop produces
$O(m_\pi^0)$ and $O(m_\pi^2)$ terms, both of which are (infinitely) large.
{\it This is not a violation of power counting}, because there are two
low-energy constants: the nucleon mass in the chiral limit, $M^{(0)}$,
and $c_{1N}$, which enter at order  $O(m_\pi^0)$ and $O(m_\pi^2)$, 
respectively, and {\it renormalize away} 
the large contributions coming from the loop.    
The renormalized relativistic result, up to and including
$O(m_\pi^3)$,  can be written as~\cite{PV05self}:
\begin{eqnarray}
M_N &=& M_N^{(0)} - 4 \, c_{1 N} \, m_\pi^2 \nonumber \\ 
&-& \frac{3 \, g_A^2}{(8 \pi f_\pi)^2} \, m_\pi^3 \, 
\left\{ 4  \left( 1 - \frac{m_\pi^2}{4 M_N^2} \right)^{5/2} 
\arccos\frac{m_\pi}{2 M_N} 
+ \frac{17 m_\pi}{16 M_N}  
- \left(\frac{m_\pi}{2 M_N}\right)^3 \right. \nonumber \\ 
&&\hspace{2.5cm} \left. + \, \frac{m_\pi}{8 M_N} 
\left[ 30 - 10 \left(\frac{m_\pi}{M_N}\right)^2 
+  \left(\frac{m_\pi}{M_N}\right)^4 \right] 
\, \ln \frac{m_\pi}{M_N}  
\right\} , 
\label{eq:nucpin}
\end{eqnarray}
and one can easily verify that the loop contribution begins
at $O(m_\pi^3)$ in agreement with power counting.

We now turn to the analysis of the
pion electroproduction process.
The pion electroproduction amplitude to NLO in the $\de$ expansion, in the
resonance region, is given by graphs in  \Figref{diagrams}(a) and (b), where the shaded
blobs in graph (a) include  corrections depicted in \Figref{diagrams}(c--f). The hadronic part of
graph (a) begins at ${\cal O}(\de^0)$ which here is the leading order. 
The Born graphs \Figref{diagrams}(b) contribute at ${\cal O}(\de)$.
We note that at NLO there are also vertex corrections of the type (e) and (f) with nucleon propagators in the loop
replaced by the $\De$-propagators. However, after the appropriate
renormalizations and $Q^2\ll\La\De$, 
these graphs start to contribute at next-next-to-leading order.

The $\Delta$ self-energy \Figref{diagrams}(c) can, in the $\zeta=\infty$
gauge, be presented as
\beq
\Si_{\al\be} (p) = \Si_\De (\slap) \,{\mathscr P}^{(3/2)}_{\al\be}(p),
\eeq
where $\Si_\De (\slap)$ has the spin-1/2 Lorentz form, 
computed by us earlier~\cite{PV05,PV05self}. To recall
these results, it is convenient to 
 introduce the following dimensionless quantities: 
\bea
\eqlab{defs}
&& \mu = \frac{m_\pi}{M_\De}, \,\,\,
r = \frac{M_N}{M_\Delta}, \,\,\, \delta =  \frac{\De}{M_\Delta} = 1-r, \nn\\ 
&& \be =\half  (1-r^2+\mu^2 ), \,\,\,
\alpha = 1 - \beta, \\ 
&& \la^2 = \quarter(\de^2-\mu^2)[(1+r)^2-\mu^2]
=\be^2-\mu^2 = \alpha^2 - r^2. \nn
\eea 
The imaginary part of the $\Delta$ self-energy is related to the
resonance decay 
width, which at the leading order (LO) gives: 
\beq
\Gamma_\Delta = -
2\, \mathrm{Im}\, \Si_\Delta(M_\Delta) = 
(4 \pi/3) \,M_\Delta \,  C_\Delta^2 \, 
\lambda^3\, (\alpha + r).
\eeq
with $C_\Delta = h_A M_\Delta / (8 \pi f_\pi)$. 
The experimental value for the 
$\De$-resonance width, $\Gamma_\Delta  = 0.115$~GeV,  
fixes $h_A\simeq 2.85$.
At NLO, the residue of the $\De$ propagator receives a correction:
\beq
{\rm Im}\,\Si_\De'(M_\De) = - 2\pi C_\De^2\,
\la \left[ \al\,\be\,  (\al+r) -\third \la^2 (r+r^2-\mu^2)\right]. 
\eeq
The renormalized $\De$-propagator, in the $\zeta=\infty$ gauge, at  
NLO reads:
\beq
S_{\al\be}(p) =
\frac{-{\mathscr P}^{(3/2)}_{\al\be}(p)}{(\slap -M_\De)[1-i\,\mbox{Im}\,\Si_\De'(M_\De)] 
+ \half i\Ga_\De} \,.
\eeq 

The real part of the renormalized $\Delta$ self-energy, 
$\tilde \Si_\Delta$,  contributes to the mass as:
\begin{eqnarray}
M_\Delta &=& M_\Delta^{(0)} - 4 \, c_{1 \Delta} \, m_\pi^2 
+ \mathrm{Re} \tilde \Si_\Delta^{(\pi N)}(M_\Delta) ,
\label{eq:delpin1}
\end{eqnarray}
where 
\begin{eqnarray}
\tilde \Si_\Delta (M_\Delta) 
= - \half C_\Delta^2 \, M_\Delta \, 
\left[ V(\mu, \delta) - V(0, \delta) - \mu^2 \, V^\prime(0, \delta) \right] .
\label{eq:delpin2}
\end{eqnarray}
The $\pi N$ loop integral $V(\mu, \delta)$ is given in analytical form 
by~\cite{PV05self}:
\bea
V(\mu, \de ) 
&=& \third (r+\al) \left[ \be \, (\mu^2-2\la^2) \,\ln\mu^2
+ \al \, (r^2-2\la^2) \,\ln r^2  -\mbox{$\frac{2}{3}$} (\al^3 + \be^3) 
\right. \nonumber \\
&+ & \left.  4\la^2
+ 4\la^4 \,\Omega(\la) \,  \right] 
+\quarter \mu^4 (\ln \mu^2 -\half ) - \quarter r^4 (\ln r^2 -\half )
\,, 
\label{eq:delpin3}
\eea
with the elementary function $\Omega$ defined as:
\beq
\Omega(\la) = \left\{ \begin{array}{lc} 
\frac{1}{2\la} 
\ln \mbox{$\frac{\be-\mu^2-\la}{\be-\mu^2+\la}$}\,, & \la^2 \ge 0\\
-\frac{1}{\sqrt{-\la^2}} 
 \arctan\frac{\sqrt{-\la^2}}{\al\be+\la^2}
\, . & \la^2 < 0 
\end{array} \right.
\eeq
The region $\lambda^2 > 0$, 
corresponds with $m_\pi < M_\Delta - M_N$, where the $\Delta$ is unstable. 
The region $\lambda^2 < 0$, 
corresponds with $m_\pi > M_\Delta - M_N$, where the $\Delta$ is stable.
  
Furthermore, from Eq.~(\ref{eq:delpin3}) we read off 
the $m_\pi^0$ and $m_\pi^2$ terms, which enter 
Eq.~(\ref{eq:delpin2}) and need to be absorbed by the renormalization of the
low-energy constants,
\begin{eqnarray}
V(0, \delta) &=& \frac{1}{4}  
\left[\frac{r}{3} (2 - r^2) + 2 (1 - \frac{r^2}{3}) \right] 
r^5 \ln r^2 
- \frac{1}{12} (1 + r)^2 (1 - r^2)^3 \ln (1 - r^2) \nonumber \\
&+& \left[ \frac{5}{36} - \frac{7 r^2}{36} - \frac{r^4}{8}  
+ \frac{r^6}{12} + \frac{r}{18} (5 - 12 r^2 + 3 r^4) \right], \\ 
V^\prime(0, \delta) &=& \frac{1}{2}  (1 + \frac{2 r}{3}) \, r^5 \ln r^2 
+ \left[ \frac{1}{3} (1 + r^2 + r^4) + \frac{r}{2} (1 + r^2) \right] 
(1 - r^2) \ln (1 - r^2) \nonumber \\
&-& \left[\frac{7}{18} + \frac{r}{2} + \frac{r^2}{6} + \frac{r^3}{2} 
+ \frac{r^4}{3} \right]. 
\label{eq:delpin5}
\end{eqnarray}

The vector-meson diagram,  \Figref{diagrams}(d), contributes to NLO for $Q^2\sim \La\De$. We include
it effectively by giving the $g_M$-term a dipole $Q^2$-dependence (in analogy to how it is usually done
for the nucleon isovector form factor): 
\beq
g_M\to \frac{g_M}{(1+Q^2/0.71\,\mbox{GeV}^2)^{2}}. 
\eeq
Analogous
effect for the $g_E$ and $g_C$ couplings begins at N$^2$LO. The loop
corrections to the $\ga N\De$-vertex are discussed separately,
in the following section.

\section{Evaluation of the $\ga N\De$ form factors}
\label{sec5}

In this section we present a detailed analysis of the $\ga N\De$-vertex 
corrections that appear at NLO, see \Figref{diagrams}(e) and (f).
To the order we work, the {\it pseudovector} $\pi NN$ interaction from
the Lagrangian \eref{Nlagran} is equivalent to the {\it pseudoscalar}
one, and we use the latter in the actual calculations of the graphs
\Figref{diagrams}(e, f).   

The rest of the section is organized as follows.
First we will comment on the electromagnetic gauge invariance
of these contributions. Then, 
after a brief summary of different decompositions
of the $\gamma N \Delta$ vertex into Lorentz-invariant form factors,
we evaluate the chiral corrections to those form factors at NLO.
As a cross-check, we do the calculations using
two different techniques: the Feynman-parameter method 
and the sideways dispersion relations.

\subsection{Electromagnetic current conservation} 

The addition of the one-loop graphs \Figref{diagrams}(e, f)
to the ``tree-level'' $\ga N\De$ vertex from \Eqref{lagran}, should not spoil the  \EM\ current conservation:
\beq
q_\mu \, \bar u_\alpha(p' ) \, \Gamma^{\alpha \mu}_{\ga N\De} \, u(p) =0\,,
\eeq
where $u_\al$ is the free Rarita-Schwinger vector-spinor of the $\De$, 
$u$ is the free Dirac spinor of the nucleon,
and momenta are defined as in \Figref{treevertex}.
This sort of conditions are most conveniently checked
by using the Ward-Takahashi (WT) identities
for the \EM\ couplings of the nucleon, pion and  $\De$ fields:
\begin{subequations}
\bea
&& q_\mu \,  \Ga_{\ga NN}^\mu = e_N\, q\cdot \ga = e_N\,[ S^{-1}_N(p') - S^{-1}_N(p)]\,,\\
&& q_\mu \,  \Ga_{\ga\pi\pi}^\mu = e_\pi\, q\cdot (p+p')= e_\pi [S^{-1}_\pi (p') - S^{-1}_\pi(p)]\,,\\
&& q_\mu \,  \Ga_{\ga\De\De}^{\mu\al\be} = e_\De\, \ga^{\mu\al\be} q_\mu
 = e_\De [S^{-1\,\al\be}_\De (p') - S^{-1\,\al\be}_\De (p)]\,,
\eea  
\end{subequations}
where $q=p'-p$, $e_N=\half e (1+\tau_3) $, $e_\pi=ie \eps^{abc} \tau_c$, $e_\De=\half e (1+3\Tau^3) $ and $S^{-1}$ denote the corresponding inverse
propagators. Applying these identities to the diagrams  \Figref{diagrams}(e, f)
we obtain:
\beq
\eqlab{cccond}
q_\mu \, \bar u_\alpha(p' ) \, \Gamma^{\alpha \mu}_{\ga N\De} \, u(p) =
e\, \bar u_\alpha(p' ) 
\,\left[\Si_{N\De}^\al(p)- \Si_{N\De}^\al (p')\right] \,u(p),
\eeq
where $\Si_{N\De}^\al$ is the one-loop $N\to \De$ self-energy, which, due
to Lorentz covariance has the following general form:
\beq
\eqlab{Lstructure}
\Si_{N\De}^\al(p) = A(p^2) \,p^\al + B(p^2) \,\ga^\al \,.
\eeq 
The scalar functions $A$ and $B$ do not vanish, however we may use
the Rarita-Schwinger conditions, $p\cdot u(p) = 0 = \ga \cdot  u(p)$,
to show that the second term on the r.h.s.\ of ~\Eqref{cccond} vanishes.
The first term can only be canceled by the contribution
of the diagram in \Figref{extra}.

Thus, the condition of
the \EM\ current conservation requires 
the diagram of \Figref{extra} to be included in addition to
the NLO contributions~\Figref{diagrams}(e, f). When computing this diagram, we
discover that it is actually of N$^2$LO, because the Lorentz form
\Eqref{Lstructure}
 cuts out the spin-3/2 part of the $\De$-propagator and hence also the
pole of the propagator. The $\De$-propagator in \Figref{extra},
therefore, counts as $\de^0$. 

To NLO, the sum of \Figref{diagrams}(e) and \Figref{diagrams}(f)
conserves the \EM\ current.

\subsection{Form-factor decompositions}

The $\gamma N \Delta$ vertex can in general be decomposed into three 
Lorentz covariants. For instance,
\begin{eqnarray}
\eqlab{newform}
\bar u_{\al}(p') \, \Gamma^{\alpha \mu}_{\ga N\De} \, u(p)   &=& 
\sqrt{\frac{3}{2}} \frac{ M_\Delta + M_N}{M_N \, [(M_\Delta + M_N)^2+Q^2]} 
\nonumber \\ 
&\times&  \bar u_{\alpha}(p^\prime) \, \left\{\, 
g_M(Q^2) \, \varepsilon^{\alpha \mu \kappa \lambda} \, 
p^{\prime}_\kappa \,  q_\lambda  \right. \nn\\
&& \hspace{1.75cm} + \; g_E(Q^2) 
\left( q^\alpha \, p^{\prime \, \mu} -q \cdot p^\prime  \, g^{\alpha \mu} \right) i \gamma_5   \\
&& \hspace{1.75cm} + \; \left.  g_C(Q^2) 
\left( q^\alpha \, q^\mu - q^2 \, g^{\alpha \mu}  \right) i \gamma_5 
\right\} \; u (p) , \nn
\label{eq:diagndel1}
\end{eqnarray}
where $q=p'-p$ is the
photon 4-momentum, $Q^2=-q^2$. 
Furthermore, $g_M(Q^2)$, $g_E(Q^2)$, and $g_C(Q^2)$ 
are the magnetic dipole, electric quadrupole, and 
Coulomb quadrupole form factors, as defined in Ref.~\cite{Caia04}.  
In the limit $Q^2=0$ they are equal to the 
physical values of the corresponding parameters 
in the Lagrangian \eref{lagran}. These
form factors relate to the conventional magnetic ($G_M^\ast$), 
electric ($G_E^\ast$) and Coulomb  ($G_C^\ast$) form factors 
of Jones and Scadron~\cite{Jones:1972ky} as follows:
\begin{eqnarray}
\eqlab{JS}
G_M^\ast &=& g_M \,+\frac{M_\De^2}{Q_+^2} \left(-\be_\ga \,g_E+  \bar Q^2 g_C\right), \nn\\ 
G_E^\ast &=& \frac{M_\De^2}{Q_+^2} \left(-\be_\ga \,g_E+ \bar Q^2  g_C\right), \\
G_C^\ast &=&-\frac{2 M_\De^2}{Q_+^2} \left(g_E +\be_\ga\,  g_C\right),\nn
\end{eqnarray}
where $Q_\pm=\sqrt{(M_\De\pm M_N)^2 +Q^2}$, $\bar Q^2 = Q^2/M_\De^2$,
$\be_\ga =\half  (1-r^2-\bar Q^2)$. 
The multipole ratios $E2/M1$ and $C2/M1$ at the $\Delta$-resonance
position can be expressed in terms of these form factors as:
\beq
\eqlab{ratios}
R_{EM}=- \frac{G_E^\ast}{G_M^\ast} \,,\;\;\;\; 
R_{SM}=- \frac{Q_+Q_-}{4M_\De^2} \, \frac{G_C^\ast}{G_M^\ast} .
\eeq

Alternatively, the on-shell $\gamma N \Delta$ vertex 
is also often expressed in the following form~:
\begin{eqnarray}
\bar u_{\al}(p') \, \Gamma^{\alpha \mu}_{\gamma N \Delta} \, u(p)  
&=&\, \mbox{$\sqrt{\frac{2}{3}}$} \, \bar u_{\alpha}(p' ) \, 
\left\{ 
\, \left( \gamma^\mu q^\alpha \, -\, \gamma \cdot q \, g^{\alpha \mu} \right) 
\, g_1(Q^2) \right. \nonumber \\
&&\hspace{2.5cm} + \left(
q \cdot p^\prime \, g^{\alpha \mu} - q^\alpha \, p^{\prime \, \mu} \right) 
\, g_2(Q^2) 
\\
&&\left. \hspace{2.5cm} 
+ \left(q^\alpha \, q^\mu - q^2 \, g^{\alpha \mu} \right) \,
g_3(Q^2) \; \right\}  i \, \gamma_5 \, u(p) ,\nn
\label{eq:diagndel0}
\end{eqnarray}
which is defined for the $p \to \Delta^+$ transition, 
yielding the isospin factor $\sqrt{2/3}$.
The form factors $g_M, g_E$ and $g_C$ can then be 
expressed in terms of $g_i(Q^2) (i = 1, 2, 3)$ 
through the relations~: 
\begin{eqnarray}
g_M(Q^2) \,&=& \, - \frac{2}{3} \, 
\frac{M_N \, Q_+^2}{M_\Delta (M_\Delta + M_N)} \, g_1(Q^2) ,
\\
g_E(Q^2) \,&=& \, \frac{2}{3} \, 
\frac{M_N \, Q_+^2}{M_\Delta (M_\Delta + M_N)} \, 
\left\{ g_1(Q^2) - M_\Delta \, g_2(Q^2) \right\} ,
\nn \\
g_C(Q^2) \,&=& \, \frac{2}{3} \, 
\frac{M_N \, Q_+^2}{(M_\Delta + M_N)} \, g_3(Q^2) .\nn
\label{eq:gmecgi}
\end{eqnarray}

\subsection{Feynman-parameter method}
The one-loop corrections to the $\gamma N \Delta$ form factors are given by 
the graphs in \Figref{diagrams}(e) and (f).
As they contain three propagators, we apply the Feynman-parameter trick~:
\beq
\frac{1}{ABC} = 2 \int_0^1 \! dx \int_0^x \!dy \frac{1}{[Ay+B(x-y)+C(1-x)]^3}.
\eeq
For the diagram  \Figref{diagrams}(e) we have  
 $A=k^2-m_\pi^2+i\eps$, $B=(k+q)^2-m_\pi^2+i\eps$, 
$C=(p-k)^2-M_N^2+i\eps$, where $k$ is the integration 4-momentum.
After the shift of the integration momentum, $k\rightarrow k-(x-y)q+(1-x)p$,
we obtain~:
$Ay+B(x-y)+C(1-x)\rightarrow k^2-\MM^2 $,
where
\beq
\MM^2(x,y) = m_\pi^2 x+ (M_N^2 -p^2 x)(1-x)- 2p\cdot q (1-x)(x-y)
-q^2 (x-y)(1-x-y) -i\eps.
\eeq

For the case of the $N \to \De$ transition we may use: 
$p^2=M_N^2$, ${p'}^2=(p+q)^2=M_\De^2$, and hence
$2p\cdot q= M_\De^2 - M_N^2 -q^2$, and 
\beq
\MM^2(x,y) = m_\pi^2 x+ [M_N^2- xM^2_\De +(M_\De^2-M_N^2) y] (1-x)
+q^2 y (x-y) -i\eps.
\eeq

The next step is to perform the integration over 4-momentum which can be done
by using the following rules:
\begin{subequations}
\bea
\eqlab{jnmunu}
&&\int \!\frac{d^4 k}{(2\pi)^4} \frac{1}{(k^2 - m^2)^{n}} \equiv J_n(m^2)
 =  i \frac{(-1)^n}{ (4 \pi)^{2} }
\frac{\Gamma (n-2)}{\Gamma (n)} \, m^{-2(n - 2)}\,,\\
&& \int \!\frac{d^4 k}{(2\pi)^4} \frac{k_\mu k_\nu}{(k^2 - m^2)^{n}}
 = \frac{1}{2(n-1)} J_{n-1}(m^2) \, g_{\mu\nu}\,, 
\eea
\end{subequations}
Integrals with an odd number of 4-vectors $k$ in the numerator vanish.
The integral $J_n$ diverges for $n=1$ and 2, but can be defined via
dimensional regularization as:
\begin{subequations}
\bea
J_1(m^2) & = & \frac{-i m^2}{(4\pi)^2} \left( 
- \frac{2}{4-d} + \ga_E -1  + \ln \frac{m^2}{4\pi}  \right), \\
J_2(m^2) & = & \frac{-i }{(4\pi)^2} \left( 
- \frac{2}{4-d} + \ga_E   + \ln \frac{m^2}{4\pi}  \right), 
\eea
\end{subequations}
where $d\to 4^-$ is the number of dimensions and $\ga_E = - \Gamma'(1) \simeq 0.5772 $
is the Euler constant.

Then, the  ($\ol {MS}$-subtracted) result for the graph of 
\Figref{diagrams}(e), where the photon couples to the charged pion in the 
loop, can be decomposed into the three Lorentz covariants \Eqref{newform}
and cast in the form~:
\begin{subequations}
\bea
g_M^{(e)} &=&  - C_{N\De} \! \int\limits_0^1 dy \, y \!\int\limits_0^{1-y} \!dx\,\ln {\cal M}_e^2 , 
\label{eq:gmpionloop} \\
g_E^{(e)} &=&  + C_{N\De}  \! \int\limits_0^1 \!dy \, y \! \int\limits_0^{1-y} \!\!dx\,
\left\{ \ln {\cal M}_e^2 
\, - 2 x \, [(1-y)\,(1+r)-x ]\,  {\cal M}_e^{-2} \right\} ,
\label{eq:gepionloop} \\
g_C^{(e)} &=&  -C_{N\De}  \! \int\limits_0^1 dy \, y \,(2 y-1)\!\!\int\limits_0^{1-y} \!\! dx 
\, [(1-y)\,(1+r)-x ]\, {\cal M}_e^{-2} , 
\label{eq:gcpionloop} \\
\mbox{with} && {\cal M}_e^2 \equiv (x-\be )^2 -\la^2 +2\be_\ga  x y + \bar Q^2 y(1-y)- i\veps ,
\eea
\end{subequations}
while the analogous contribution of the graph \Figref{diagrams}(f), where the photon 
couples to the electric charge of the nucleon in the loop, is given by~:
\begin{subequations}
\bea
g_M^{(f)} &=&  - C_{N\De} \!\int\limits_0^1 dy \, y \!\int\limits_0^{1-y} \!dx\,\left\{ 2\ln {\cal M}_f^2  \right. \nn\\
&& \left. + \,
{\cal M}_f^{-2}
\left[ 2\al_\ga  x y + \bar Q^2 y(1-y)-(x+r)(1-x)\right]\right\}, 
\label{eq:gmnucloop} \\
g_E^{(f)} &=& -g_M^{(f)} + 2 C_{N\De} \! \int\limits_0^1 \!dy \, y \! 
\int\limits_0^{1-y} \!\!dx\,{\cal M}_f^{-2}
\, \left[ (x+r)(1-x)-xy(1+r)\right] , 
\label{eq:genucloop} \\
g_C^{(f)} &=&  -2 C_{N\De}  \int\limits_0^1 dy \, y^2 
\int\limits_0^{1-y} dx\,[y(1+r)-1+x]\,  {\cal M}_f^{-2},
\label{eq:gcnucloop} \\
\mbox{with} && {\cal M}_f^2  
\equiv  (x-\al)^2 -\la^2 +2\al_\ga  x y + \bar Q^2 y(1-y)- i\veps,
\eea
\end{subequations}
where we have used the definitions of \Eqref{defs} and introduced,
\begin{equation}
C_{N\De} = \frac{4 g_A h_A}{(8 \pi f_\pi)^2} 
\frac{Q_+^2 r^2}{3 \, (1+r)}. 
\end{equation}

Finally, the integration over $x$ and some of the $y$-integration can be
done analytically, see Appendix A.

\subsection{Dispersion method}

Alternatively, we can  compute 
the correction to the $\gamma N \Delta$ vertex 
by exploiting the {\it analyticity} of the loop
contributions. We use the so-called
{\it sideways dispersion relations}~\cite{bincer60,HPV05},
and hence start with calculating 
the absorptive (imaginary) part of the $\gamma N \Delta$ vertex 
from the cut in the $\pi N$-loop diagrams of Fig.~\ref{fig:gandelabs}. 
Subsequently, the real part of the on-shell 
$\gamma N \Delta$ vertex is computed through the dispersion relation 
in the $\Delta$-resonance 4-momentum squared ${p'}^2$. 

We first express the half off-shell $\gamma N \Delta$ 
vertex (where we allow the $\Delta$ to be off-shell) in terms of 
the invariants $g_i(s = {p'}^2, Q^2)$, for $i = 1, 2, 3$, as~:
\begin{eqnarray}
\Gamma^{\alpha \mu}  \, u(p)  
&=&\, \left({\cal P}^{3/2}_+ (p^\prime)\right)^\alpha_{\; \; \beta} \, 
\left\{\, 
\, \left( \gamma^\mu q^\beta \, -\, \gamma \cdot q \, g^{\beta \mu} \right) 
\, g_1^+(s,\, Q^2) \right. \nonumber \\
&&\hspace{2.5cm} + \left(
q \cdot p^\prime \, g^{\beta \mu} - q^\beta \, p^{\prime \, \mu} \right) 
\, g_2^+(s,\, Q^2) 
\nonumber \\
&&\left. \hspace{2.5cm} 
+ \left(q^\beta \, q^\mu - q^2 \, g^{\beta \mu} \right) \,
g_3^+(s,\, Q^2) \; \right\}  i \, \gamma_5 \, u (p) \nonumber \\
&+& \left({\cal P}^{3/2}_- (p^\prime)\right)^\alpha_{\; \; \beta}  \, 
\left\{\, 
\, \left( \gamma^\mu q^\beta \, -\, \gamma \cdot q \, g^{\beta \mu} \right) 
\, g_1^-(s,\, Q^2) \right. \nonumber \\
&&\hspace{2.5cm} + \left(
q \cdot p^\prime \, g^{\beta \mu} - q^\beta \, p^{\prime \, \mu} \right) 
\, g_2^-(s,\, Q^2) 
\nonumber \\
&&\left. \hspace{2.5cm} 
+ \left(q^\beta \, q^\mu - q^2 \, g^{\beta \mu} \right) \,
g_3^-(s, \,Q^2) \; \right\}  i \, \gamma_5 \, u(p) ,
\label{eq:regndel}
\end{eqnarray}
where the spin-3/2 positive (negative) energy projectors are given by 
\begin{eqnarray}
\left( {\cal P}^{3/2}_\pm (p^\prime) \right)_{\alpha \beta} = 
\frac{(\gamma \cdot p^\prime \pm M_\Delta)}{2 \, M_\Delta} \, 
\left\{ g_{\alpha \beta} - \frac{1}{3} \gamma_\alpha \gamma_\beta 
- \frac{1}{3 \, {p'}^2 } 
\left( \gamma \cdot p^\prime \gamma_\alpha p^{\prime}_{\beta} 
+ p^{\prime}_\alpha \gamma_\beta \gamma \cdot p^\prime  \right) 
\right\}.
\end{eqnarray}

We next determine the real parts of the invariants $g_i^+$  
for $s = M_\Delta^2$ (i.e., the $\Delta$ is on-shell) 
through a dispersion relation in $s$ as~:
\begin{eqnarray}
\mathrm{Re} \, g_i^+ (M_\Delta^2, Q^2) = \frac{1}{\pi} \,
{\mathrm P} \!\int\limits_{s_{th}}^\infty \! d s \, 
\frac{\mathrm{Im} \, g_i^+ (s, Q^2)}{s - M_\Delta^2},
\end{eqnarray}
where the integration starts from the $\pi N$ threshold
$s_{th} =(M_N + m_\pi)^2$, and 
where ${\mathrm P}$ denotes the principal value integration. 
These unsubstracted dispersion integrals do not converge. However,
the dispersion relations for the following differences
\begin{eqnarray}
\eqlab{subdr}
\mathrm{Re} \, g_i^+ (M_\Delta^2, Q^2) - \mathrm{Re} \, g_i^+ (M_\Delta^2, 0) 
= \frac{1}{\pi} \,
{\mathrm P}\!\int\limits_{s_{th}}^\infty \!
 d s \, 
\frac{\mathrm{Im} \, g_i^+ (s, Q^2) 
- \mathrm{Im} \, g_i^+ (s, 0)}{s - M_\Delta^2},
\label{eq:subtdr} 
\end{eqnarray}
do converge, for all the considered contributions. This subtraction can be put in correspondence with the
renormalization of appropriate low-energy constants from the Lagrangian~\eref{lagran}. 

The expressions for the 
absorptive (imaginary) parts of these form factors are calculated
in Appendix \ref{sec:phase} with the results given 
by Eqs.~\eref{gipi}, \eref{giN}.
The dispersion integrals in \Eqref{subdr} are evaluated numerically.
The physical form factors are then simply given by
$g_i(Q^2) = g_i^+(M_\Delta^2, Q^2)$. 

We compared the thus obtained results with 
the  Feynman-parameter method results.  
We found perfect agreement between both methods 
for all three $\gamma N \Delta$ transition form factors, which 
provides a cross-check on our calculations.

\section{Errors due to neglect of N$^2$LO effects}
\label{sec4}

Prior to presenting the results of our NLO calculation,
we would like to make an estimate of the theoretical uncertainty
due to the neglect of higher-order effects. Some of the
next-next-to-leading order (N$^2$LO) 
contributions are shown in \Figref{NNLOgraphs}. Of course,
there is no substitute for an actual calculation of those
effects, but at present we only know that they must be suppressed
by at least one power of $\de$ ($=\De/\La_{\chi SB}$) 
as compared to the NLO and two powers of $\de$  
as compared to the LO contributions. Therefore, we can {\it estimate}
the size of the N$^2$LO contribution to an amplitude $A$ as:
$A_{\mathrm{NLO}}\, \de $, or $A_{\mathrm{LO}}\, \de^2 $. The theoretical
uncertainty of a calculation up to and including NLO can thus be
estimated as:
\begin{subequations}
\eqlab{naive}
\beq
A_{err} = | A_{\mathrm{LO}} + A_{\mathrm{NLO}}| \, \de^2.
\eeq
In cases where the amplitude does not receive any LO contributions,
we have no other option than,
\beq
A_{err} = | A_{\mathrm{NLO}}| \, \de.
\eeq
\end{subequations}

This looks nice and simple, however, there are a few caveats in the
implementation of such an estimate. First of all, although we have
introduced $\de$ as $\De/\La_{\chi SB}$, it counts as well
the other small scales of the theory, $m_\pi$ and $Q^2$.
Therefore, we shall estimate the error using the following
expansion parameter (assuming $\La_{\chi SB}\sim M_N$): 
\beq
\tilde \de = \frac{1}{3}
 \left[\frac{\De}{M_N} + 
\left(\frac{m_\pi}{M_N}\right)^{1/2} 
+ \left(\frac{Q^2}{M_N^2}\right)^{1/2} \right],
\eeq
where all the light scales are treated on equal footing and 
hence are averaged over. 

Secondly, what if the amplitude happens to vanish at some
kinematical point. According to \Eqref{naive} the theoretical calculation
at that point would be perfect, which is of course  unlikely to be true
in reality. So, when considering dependencies on kinematical variable(s),
we shall take an {\it average of the error} over some appropriate
region of that variable.  

Given these two points, we are led to the following 
formula for the theoretical uncertainty of the NLO
calculation for an amplitude $A$,
\beq
\eqlab{Aerror}
A_{err} = \left\{ \begin{array}{cc}
|A|_{av}  \, \tilde\de^2, & \mbox{LO}\neq 0 \\
|A|_{av}  \, \tilde\de, & \mbox{LO} = 0,\,
\end{array}\right.
\eeq
and the subscript ``{\it av}'' indicates
that the appropriate averaging is performed.

The theoretical uncertainty of the NLO
calculation
of an observable $O$ is:
\beq
\eqlab{Oerror}
O_{err} = \left\{ \begin{array}{cc}
2 |O|_{av}  \, \tilde\de^2, & \mbox{LO}\neq 0 \\
2 |O|_{av}  \, \tilde\de, & \mbox{LO} = 0,\,
\end{array}\right.
\eeq
where the factor of 2 takes into account that an observable
is a product of two amplitudes.

Note that this error estimate differs from the one presented 
in our earlier paper~\cite{Pascalutsa:2005ts}, where we overlooked the case
of vanishing LO contributions. This underestimated the
error in some observables, as will be discussed in more detail below.

\section{Results and discussion}
\label{sec6}

We are now in position to discuss the NLO results for 
pion photo- and electroproduction amplitudes and observables. 
We begin with the multipole-analysis for pion photoproduction. 
The resonant photoproduction multipoles are well-established,
and we use them to determine the two low-energy constants:
$g_M$ and $g_E$, the strength of the $M1$ and $E2$ 
$\gamma N \Delta$ transitions. 
Subsequently, we discuss the results for pion 
{\it electro}production: cross sections, 
multipoles and the $Q^2$ dependence of the $R_{EM}$ and $R_{SM}$ ratios. 
The comparison with pion electroproduction observables allows to determine 
the third low energy constant in the \ceft\ framework, related to the 
strength of the $C2$ $\gamma N \Delta$ transition.
Once the three low energy constants are determined, 
the $Q^2$- and $m_\pi$-dependencies follow as a prediction of the NLO result. 
We discuss the  predictions for the $m_\pi$-dependence of the magnetic 
dipole $\gamma N \Delta$ form factor, the $R_{EM}$ and $R_{SM}$ ratios, 
and compare with the recent lattice-QCD results.

\subsection{Pion photoproduction}

In Fig.~\ref{fig:gap_pin_mult}, we show the result of the \ceft\  calculations 
for the pion photoproduction resonant multipoles 
$M_{1+}^{(3/2)}$ and $E_{1+}^{(3/2)}$, around the 
resonance position, as function of the total {\it c.m.} energy $W$ 
of the $\pi N$ system. 
These two multipoles are well established by the MAID~\cite{Drechsel:1998hk}
and SAID~\cite{Arndt:2002xv} partial-wave solutions which allow us to fit the two low-energy constants 
of the chiral Lagrangian 
\Eqref{lagran} as~:~$g_M = 2.9$, $g_E = -1.0$.   As is seen from the figure,
with these values the NLO results (solid lines) give a good description of the energy 
dependence of the resonant multipoles in 
a window of 100 MeV around the $\Delta$-resonance position.
Also, these values yield $R_{EM}= -2.3$ \%, 
in a nice agreement with experiment~\cite{Mainz97}.

The dashed curves in Fig.~\ref{fig:gap_pin_mult} 
show the contribution of the $\Delta$-resonant diagram of \Figref{diagrams}(a)
{\it without} the NLO vertex corrections \Figref{diagrams}(e, f).
For the $M_{1+}$ multipole this is the LO contribution.
For the $E_{1+}$ multipole
the LO contribution is absent [recall that the $g_E$ coupling
is of one order higher than $g_M$, see \Eqref{lagran}]. 
Hence,  the dashed curve represents
a partial NLO contribution to $E_{1+}$ therein.

Upon adding the non-resonant  Born graphs \Figref{diagrams}(b) to the
dashed curves we obtain the dotted curves in 
Fig.~\ref{fig:gap_pin_mult}. These non-resonant contributions are purely 
real at this order and do not affect the imaginary part of the multipoles. 
One sees that the resulting calculation is flawed because the real 
parts of the resonant multipoles now fail to cross zero at the resonance 
position and hence unitarity, in the sense of Watson's theorem~\cite{Wat54},
is violated.\footnote{Recall that Watson's
theorem is a simple statement of unitarity based on a coupled-channel
scattering equation to leading order in the electromagnetic 
interaction. The theorem relates the phase of a
photoproduction multipole $M_l^{(\ga N)}$ 
to a corresponding (in spin and parity)
$\pi N$-scattering phase-shift $\de^{(\pi N)}_l$:
$$
M_l^{(\ga N)} = | M_l^{(\ga N)} | \, e^{i\de^{(\pi N)}_l}\,.
$$
For the resonant channel the corresponding phase-shift 
crosses 90 degrees
at the resonance position, and thus the real
part of the resonant multipoles must vanish.}
The complete NLO calculation, shown by the solid curves in the figure
includes in addition the vertex 
corrections \Figref{diagrams}(e, f), which restore unitarity 
{\it exactly}. This may come as a surprise, since we are doing a perturbative
calculation, without a resummation of rescattering contributions. 
Nevertheless, it not difficult to see that our 
numerical result is not a fluke. {\it Watson's theorem is satisfied exactly by
the NLO, up to-one-loop  amplitude} 
given the graphs in \Figref{diagrams}.

It is also interesting to examine our calculation for the non-resonant
multipoles, which all receive contributions of NLO only. 
In Fig.~\ref{fig:gap_pin_nonresmult}, we show the NLO calculations for 
the real parts of the non-resonant $s$-, $p$- and $d$-wave 
pion photoproduction multipoles in the $\Delta(1232)$ region in 
comparison with the 
two state-of-the-art phenomenological multipole 
solutions, MAID and SAID.  
Note that at NLO in the $\delta$-expansion, the non-resonant 
multipoles are purely real. The multipole solutions show 
indeed that the imaginary 
parts of non-resonant multipoles, around the $\Delta$ resonance,
are negligibly smaller than their real parts. 
From the figure one thus sees that for most of the non-resonant 
multipoles, the parameter-free NLO results, 
agree fairly well with the phenomenological 
multipole solutions.  
The largest differences are observed for $M_{1-}^{(1/2)p}$ multipole.   
This multipole corresponds with nucleon quantum numbers. The cause 
of the appreciable difference in this channel is largely 
due to the nucleon anomalous-magnetic-moment
contributions, which are not included in our calculation (since
they appear at N$^2$LO in the 
$\delta$-expansion), but which are included in the 
phenomenological solutions.

\subsection{Pion electroproduction}

We now  will present the electroproduction observables obtained from
the NLO amplitude of \Figref{diagrams}.
The five-fold pion electroproduction cross section can be 
expressed as~:
\begin{eqnarray}
\frac{d \sigma}{(d E_e^\prime \, d \Omega_e^\prime)^{lab} \, 
d \Omega_\pi^{c.m.}} \,=\, \Gamma_v \, \frac{d \sigma}{d \Omega_\pi^{c.m.}},
\end{eqnarray}
where the virtual photon flux factor $\Gamma_v$ is defined as~:
\begin{eqnarray}
\Gamma_v = \frac{e^2}{(2 \pi)^3}\, \frac{E_e^\prime}{E_e}\,
\frac{(W^2 - M_N^2)}{2 M_N} \,
\frac{1}{ (1 - \varepsilon)\,Q^2}, 
\end{eqnarray}
where  $W$ is  the invariant mass of the final $\pi N$ system,
$E_e$ (${E_e}'$) are the initial (final) electron {\it lab} 
energies, and $\varepsilon$ denotes the photon polarization parameter. 

The $\gamma^* N \to \pi N$ cross section 
for unpolarized nucleons is expressed in terms of 5 response functions 
as~:
\begin{eqnarray}
\eqlab{xsecn}
\frac{d \sigma}{d \Omega_\pi} &=&
\frac{d \sigma_T}{d \Omega_\pi} 
+ \varepsilon \, \frac{d \sigma_L}{d \Omega_\pi} + \varepsilon \, \cos 2 \Phi \, 
\frac{d \sigma_{TT}}{d \Omega_\pi} \nn \\
&+&
\sqrt{2 \varepsilon  (1 + \varepsilon) }\, \cos \Phi \, 
\frac{d \sigma_{LT}}{d \Omega_\pi}  
+ h \sqrt{2 \varepsilon  (1 - \varepsilon) }\, \sin \Phi \, 
\frac{d \sigma_{LT}^ \prime}{d \Omega_\pi} , 
\end{eqnarray}
where $\Th_\pi$ and $\Phi$ are the pion polar and azimuthal 
{\it c.m.} angles, respectively, 
and $h$ denotes the electron helicity.

In \Figref{crossections} we show the NLO results for the 
different virtual photon absorption cross sections 
entering \Eqref{xsecn} at the resonance position, and 
for $Q^2 \simeq 0.127$~GeV$^2$, where recent precision data are available. 
Besides the low-energy constants $g_M$ and $g_E$, which were fixed 
from the resonant multipoles in Fig.~\ref{fig:gap_pin_mult}, 
the only other low-energy constant from \Eqref{lagran} 
entering the NLO electroproduction calculation is $g_C$. 
In \Eqref{xsecn}, the main sensitivity on $g_C$ enters in 
$\sigma_{LT}$. A best description of the $\si_{LT}$ data in 
\Figref{crossections} is obtained by choosing 
$g_C = -2.36$. 

The theoretical uncertainty
of the NLO result is estimated by using \Eqref{Oerror}, where 
the average is taken over the range of  $\Theta_\pi$.
Note that $\si_{LT}$ and $\si_{LT'}$ do not receive any LO
contributions and therefore the LO$=0$ case in \Eqref{Oerror}
must be applied in the estimate. This point was overlooked
in our first error estimates of the NLO calculation~\cite{Pascalutsa:2005ts},
which clearly led to an underestimate of the theory error
for  $\si_{LT}$ and $\si_{LT'}$.

From \Figref{crossections}, one 
sees that the NLO \ceft\ calculation, within its accuracy,
is consistent with the experimental data for these observables.

The reliability of the present \ceft\ calculation 
can also be tested by comparing its predictions for the non-resonant 
multipoles with the phenomenological multipole solutions MAID and SAID. In 
\Figref{ep_epin_nonresmult}, we show this comparison for the 
non-resonant $s$- and $p$-wave pion electroproduction 
multipoles at the resonance position as function of $Q^2$. 
Note that there is 
a considerable uncertainty in the phenomenologically extracted  
$s$-wave scalar multipoles $S_{0+}^{(3/2)}$ and $S_{0+}^{(1/2)p}$ at low $Q^2$. 
As one can see from the figure, our calculation is in a reasonable
agreement with the phenomenological solutions for 
most of the $s$- and $p$-wave non-resonant multipoles.
 The largest discrepancy is observed in the
$Q^2$-dependence of both $E_{0+}^{(3/2)}$ and $E_{0+}^{(1/2)p}$. 
This discrepancy will need to be resolved by
$s$-wave N$^2$LO corrections which grow with $Q^2$.

In \Figref{ratiosQ2} we show the $Q^2$ dependence of the ratios $R_{EM}$ and $R_{SM}$. Having 
fixed the low energy constants $g_M$, $g_E$ and $g_C$, this $Q^2$ dependence follows as a prediction.  
The theoretical uncertainty here (shown by  
the error bands) is estimated according to \Eqref{Aerror}, for LO=$0$
case, and the average 
taken over the range  of $Q^2$ from 0 to 0.2~GeV$^2$. 
From the figure one sees that the  
NLO calculations are consistent with the experimental data for both
of the ratios.

To see how higher-order effects may affect the $Q^2$ dependence, 
we include the vector-meson type of dependence for the electric
$\ga N\De$ transition by the following replacement of the 
low-energy constant:
\beq
\eqlab{geqsqr}
g_E \to \frac{g_E}{\left(1 + Q^2 / \Lambda_E^2 \right)^2} 
\eeq 
In contrast to the analogous effect for $g_M$, which is of NLO,
the inclusion of the $Q^2$ dependence in $g_E$ is a N$^2$LO effect. 
We choose $\Lambda_E$ so as to satisfy the asymptotic condition~\cite{pQCD}: $R_{EM} \to 1$, 
for $Q^2 \to \infty$. This is achieved by 
the choice~: $\Lambda_E^2 = \sqrt{ - g_M(0) / g_E(0) } \, 0.71$GeV$^2$.   
The dashed curve in \Figref{ratiosQ2} shows the resulting
effect of the replacement \eref{geqsqr}, as imposed on the
complete NLO result shown by the solid curves with error bands.
The fact that the dashed curves go out of the error bands at some point
indicates that our error estimate is not designed for such high $Q^2$ values.

\subsection{Chiral behavior and chiral extrapolations }

Since the low-energy constants $g_M$, $g_E$, and $g_C$ have been fixed, our  
calculation can provide a prediction for the $m_\pi$ dependence of the 
$\gamma N \Delta$ transition form factors. The study of the $m_\pi$-dependence 
is crucial to connect to lattice QCD results, which at present 
can only be obtained for larger pion masses (typically $m_\pi \gtrsim 300$ MeV). 

The $m_\pi$ dependence of the nucleon and $\De$-resonance
masses, given above by  Eq.~(\ref{eq:nucpin}) and (\ref{eq:delpin2}),
are compared with lattice results in 
\Figref{nucdelmass}. We constrain one of the two parameters
in Eq.~(\ref{eq:nucpin}) 
by the physical nucleon mass value at $m_\pi = 0.139$ GeV,  
while the other parameter
is fit to the lattice data shown in the figure. 
This yields~: $M_N^{(0)} = 0.883$~GeV and $c_{1N} = -0.87$~GeV$^{-1}$. 
As is seen from the figure, with this two-parameter form for $M_N$, 
a good description of lattice results is obtained 
up to $m_\pi^2 \simeq 0.5$~GeV$^2$.

Analogously to the nucleon case, 
we fix one parameter in Eq.~(\ref{eq:delpin2}) 
from the physical value of the $\Delta$ mass,
while the second parameter is fit to the lattice data
shown in \Figref{nucdelmass},
yielding~: $M_\Delta^{(0)} = 1.20$~GeV and 
$c_{1 \Delta} = -0.40$~GeV$^{-1}$. 
As well as for the nucleon, 
this two-parameter form for $M_\Delta$ yields a fairly good description 
of the lattice results up to $m_\pi^2 \simeq 0.5$~GeV$^2$.

In Fig.~\ref{fig:regmmpi} we examine the $m_\pi$-dependence of the 
{\it magnetic} $\gamma N \Delta$-transition form factor $G_M^\ast$, in  
the convention of Jones and Scadron, see \Eqref{JS}. 
At the physical pion mass, this form 
factor can be obtained from the imaginary part of the $M_{1+}^{3/2}$ multipole 
at $W = M_\Delta$ (where the real part is zero by Watson's theorem) as~:
\begin{eqnarray}
{\rm Re} \, G_M^\ast (Q^2) = 
\left( \frac{8 \, M_N^2 \, |{\bf p}_\pi^\ast| \, \Gamma_\Delta }{3 \, \alpha_{em} \, |{\bf q}^\ast|^2}  \right)^{1/2} \,
{\rm Im} \, M_{1+}^{3/2} (W = M_\Delta, Q^2), 
\end{eqnarray}
where $|{\bf p}_\pi^\ast|$ ($|{\bf q}_\pi^\ast|$) denote the 
pion (virtual photon) {\it c.m.} three-momenta respectively 
at the resonance position, i.e. for $W = M_\Delta$.  
Recall that the value of $G_M^\ast$ at $Q^2 = 0$ is 
determined by the low-energy constant $g_M$.
The $Q^2$-dependence then follows as a prediction of the NLO
result, and Fig.~\ref{fig:regmmpi} shows that
this prediction is consistent with the experimental 
value at $Q^2 = 0.127$~GeV$^2$ and physical pion mass. 

The $m_\pi$-dependence  
of $G_M^\ast$ is also completely fixed at NLO, no new parameters appear. 
In Fig.~\ref{fig:regmmpi}, the result for $G_M^\ast$ at
$Q^2 = 0.127$~GeV$^2$ is shown both when the $m_\pi$-dependence of 
the nucleon and $\Delta$ masses is included and when it is not.
Accounting for the $m_\pi$-dependence in $M_N$ and $M_\Delta$, 
shown in Fig.~\ref{fig:nucdelmass}, apparently   
changes the result for $G_M^\ast$ quite significantly. 
The \ceft\  calculation,  
with the $m_\pi$ dependence of $M_\N$ and $M_\Delta$ included, is in
a qualitatively good agreement with the lattice data shown in the figure. 
The \ceft\ result also follows 
an approximately linear behavior in $m_\pi^2$, 
although it falls about 10 - 15 \% below the lattice data.  
This is just within the uncertainty of the NLO results.
One should also keep in mind that the present lattice simulations
are not done in full QCD, but are ``quenched'', so discrepancies
are not unexpected.

In \Figref{ratios}, we show the $m_\pi$-dependence of the ratios 
$R_{EM}$ and $R_{SM}$ and compare them to lattice QCD calculations.  
The recent state-of-the-art lattice calculations of 
$R_{EM}$ and $R_{SM}$~\cite{Ale05} use a {\it linear}, 
in the quark mass ($m_q\propto m_\pi^2$), {\it extrapolation}
to the physical point,  
thus assuming that the non-analytic $m_q$-dependencies are  negligible. 
The thus obtained value for $R_{SM}$ at the physical 
$m_\pi$ value displays a large 
discrepancy with the  experimental result, as seen in \Figref{ratios}. 
Our calculation, on the other hand, shows  that the non-analytic dependencies 
are {\it not} negligible. While
at larger values of $m_\pi$, 
where the $\Delta$ is stable, the ratios display a smooth 
$m_\pi$ dependence, at $m_\pi =\De $ there is an inflection point, and 
for  $m_\pi \leq \Delta$ the non-analytic effects are crucial, 
as was also observed for the $\De$-resonance
magnetic moment~\cite{Cloet03,PV05}.

One also sees from \Figref{ratios} that, unlike the result for 
$G_M^\ast$, there is only little difference between the \ceft\ 
calculations with the $m_\pi$-dependence of 
$M_N$ and $M_\Delta$ accounted for, and our earlier calculation
\cite{Pascalutsa:2005ts}, 
where the ratios were evaluated neglecting the $m_\pi$-dependence of the masses. 
This is easily understood, as the main effect due to the $m_\pi$-dependence 
of $M_N$ and $M_\Delta$ arises due to a common factor in the evaluation 
of the $\gamma N \Delta$ form factors, which drops out of the ratios.
One can speculate that the ``quenching'' effects 
drop out, at least partially, from the ratios as well.

In \Figref{ratios} we also show the $m_\pi$-dependence of the $\gamma N \De$ transition ratios, 
with the theoretical uncertainty  estimated according to \Eqref{Aerror},
for the case LO$=0$, and with 
the average
taken over the range  of $m_\pi^2 $ from 0 to 0.15~GeV$^2$. 
The $m_\pi$ dependence obtained here from \ceft\  clearly shows that
the lattice results for $R_{SM}$ may in fact be consistent with experiment.

\section{Conclusion}
\label{sec7}

Let us briefly go over the main points and results presented in this paper,
which is the first one in a series devoted to \ceft\ in the
$\De$(1232)-resonance region.
\begin{itemize}
\item[(i)] 
We  develop an extension of chiral perturbation theory to the
$\Delta$(1232)-resonance energy region, based
on the $\de$-expansion of Ref.~\cite{PP03}.
In this \ceft\ framework the expansion is done in 
the small parameter $\de$ equal to the excitation
energy of the resonance over the chiral symmetry-breaking scale.
The other low-energy scale of the theory, the pion mass,
counts as $\de^2$, which is a crucial point for an adequate counting
of the $\De$-resonance contributions in both the low-energy
and the resonance energy regions.  
\item[(ii)]
This framework has been applied here to the process of
pion electroproduction. 
This is a first \ceft\ study of this reaction
in the $\Delta(1232)$-resonance region.
We have performed a complete calculation of this process in the
resonance region up to,
and including, next-to-leading order in the  $\de$-expansion.
The power counting in $\de$ has only been used to establish which graphs 
contribute at the leading and next-to-leading order, 
no actual expansion of the diagrams themselves is being done. 
Therefore, some higher-order in $\de$ effects, required
by relativity and analyticity, are automatically included. 
Such effects are known to
improve the convergence and 
extend the region of applicability of \ceft\ calculations.
\item[(iii)] Our NLO calculation of pion electroproduction
satisfies {\it gauge and chiral symmetries} perturbatively, and
{\it Lorentz-covariance, analyticity, unitarity (Watson's theorem)} exactly.
\item[(iv)] The chiral-loop contributions to the
$\gamma N \Delta$ transition have been evaluated 
using two independent techniques: the 
Feynman parameter method and  the 
sideways dispersion relations. 
Both methods yield the same result. 
\item[(v)] The only free parameters entering at this order
are the $\ga N\De$ couplings $g_M$, $g_E$, $g_C$ characterizing
the $M1$, $E2$,  $C2$ transitions, respectively. 
By comparing our NLO results with the standard 
multipole solutions (MAID and SAID) for the photoproduction multipoles
we have extracted $g_M$ = 2.9 and 
$g_E = -1.0$, corresponding to $R_{EM} = -2.3$~\%.  
The NLO \ceft\ 
result was also found to give a good description of the 
energy-dependence of most non-resonant $s$, $p$ and $d$-wave 
photoproduction multipoles in a 100 MeV window 
around the $\Delta$-resonance position. 
From the pion electroproduction cross-section $\si_{LT}$
we have extracted $g_C = -2.36$, which yields $R_{SM} \simeq -7$~\% 
near $Q^2 = 0.1$~GeV$^2$. In overall, the NLO
results are consistent with the experimental data
of the recent high-precision measurements at MAMI and BATES.  
\item[(vi)]
The \ceft\ framework plays a {\it dual role}
in that it allows for
an extraction of resonance parameters from observables {\em and} 
predicts their pion-mass dependence. In this way it may provide
 a crucial connection of present lattice QCD results (obtained 
at unphysical values of $m_\pi$) to the experiment.
We have shown here that the opening of the 
$\De\to \pi N$ decay channel at $m_\pi = M_\De-M_N$
induces a pronounced  non-analytic
behavior of the $R_{EM}$ and $R_{SM}$ ratios. 
While the linearly-extrapolated lattice QCD results 
for $R_{SM}$ are in disagreement with experimental data, the 
\ceft\ prediction of the non-analytic dependencies suggests that
these results are in fact consistent with experiment.
\item[(vii)]
The present calculation is systematically improvable. 
We have indicated what are the next-next-to-leading order effects,
however,  at present we could 
only estimate the theoretical uncertainty of our calculations
due to such effects. We have defined and provided a corresponding
error band on our NLO results. An actual calculation of 
N$^2$LO effects is a worthwhile topic for a future work.  
\end{itemize}

As high-precision data for low-$Q^2$ pion electroproduction 
in the $\Delta$-resonance region become available 
from BATES, MAMI and JLab, 
and the  next-generation lattice calculations of the 
$\gamma N \Delta$ transition are on the way~\cite{Alexandrou:2005em}, 
the \ceft\ presented here makes a promise 
to be the theoretical framework to 
examine and connect these results.

\appendix
\section{Further evaluation of the Feynman-parameter integrals}
\label{app1}

To perform the integrals in Eqs.~(\ref{eq:gmpionloop}-\ref{eq:gcpionloop}), 
we first note that ${\cal M}_e^2 $
can be written as $[(x-\be + y\be_\ga)^2 - {\cal D}^2]$, with
\beq
{\cal D}(y) = \left[ (\be -y \be_\ga  )^2 -\mu^2 - \bar Q^2 y (1-y)+i\veps \right]^{1/2} \,,
\eeq
and then use the following elementary integrals
\bea
&& \int dx \, \ln [(x-a)^2 -b^2] = -2x +(x-a)\,\ln[(x-a)^2 -b^2] + b\,\ln \frac{x-a+b}{x-a-b} + C, \nn\\
&& \int dx \, \frac{1}{(x-a)^2 -b^2} = \frac{1}{2b} \,\ln \frac{x-a+b}{x-a-b} + C, \nn\\
&& \int dx \, \frac{x}{(x-a)^2 -b^2} = \half \ln[(x-a)^2 -b^2]+ \frac{a}{2b} \,\ln \frac{x-a+b}{x-a-b} + C, \\
&& \int dx \, \frac{x^2}{(x-a)^2 -b^2} = x+ a\, \ln[(x-a)^2 -b^2]+ \frac{a^2+b^2}{2b} \,\ln \frac{x-a+b}{x-a-b} + C, \nn
\eea
to perform one integration:
\bea
g_M^{(e)} &=&  - C_{N\De} \! \int\limits_0^1 dy \, y \!\int\limits_0^{1-y} \!dx\,\ln {\cal M}_e^2 
= - C_{N\De} \! \int\limits_0^1 dy \, y \, \left\{ -2(1-y) \right. \nn\\
& + &  (\al - y\al_\ga) \ln
[ r^2(1-y)^2 + y \mu^2] + (\be - y\be_\ga) \ln [ \mu^2 + \bar Q^2 y(1-y)]  \nn\\
&+& \left. {\cal D} \,
\ln \left[ \frac{(1-y)(  \be - y\be_\ga - y \bar Q^2 + {\cal D})  - \mu^2 +i\veps }{(1-y)
(  \be - y\be_\ga - y \bar Q^2 - {\cal D})  - \mu^2 +i\veps }\right] 
\right\}, \nn\\
g_E^{(e)} &=& C_{N\De}  \! \int\limits_0^1 \!dy \, y \! \int\limits_0^{1-y} \!\!dx\,
\left\{ \ln {\cal M}_e^2  - 2 x \, [-x +(1-y)\,(1+r) ]\,  {\cal M}_e^{-2} \right\} =
- C_{N\De}  \! \int\limits_0^1 \!dy \, y \nn \\
&\times & \left\{ (\be - y\be_\ga - r(1-y)\,) \, \ln[ r^2(1-y)^2 + y \mu^2]
+ (\al - y\al_\ga + r(1-y)\,)\ln [ \mu^2 + \bar Q^2 y(1-y)]  \right. \nn \\
&+& \left. \frac{\be-y \be_\ga}{{\cal D}} (\al - y\al_\ga + r(1-y)\,) \,
\ln \left[ \frac{(1-y)(  \be - y\be_\ga - y \bar Q^2 + {\cal D})  - \mu^2 +i\veps }{(1-y)
(  \be - y\be_\ga - y \bar Q^2 - {\cal D})  - \mu^2 +i\veps }\right] 
\right\}, \\
g_C^{(e)} &=&  -C_{N\De}  \! \int\limits_0^1 dy \, y \,(2 y-1)\!\!\int\limits_0^{1-y} \!\! dx
\, [-x +(1-y)\,(1+r) ]\, {\cal M}_e^{-2} \nn \\
&=&  -C_{N\De}  \! \int\limits_0^1 dy \, y \,(2 y-1) \,
 \half \left\{ -\ln\frac{ r^2(1-y)^2 + y \mu^2}{\mu^2 + \bar Q^2 y(1-y)} \right. \nn\\
&& \left. \hspace{2cm}
- \frac{1}{{\cal D}} (\al - y\al_\ga + r(1-y)\,) \,
\ln \left[ \frac{(1-y)(  \be - y\be_\ga - y \bar Q^2 + {\cal D})  - \mu^2 +i\veps }{(1-y)
(  \be - y\be_\ga - y \bar Q^2 - {\cal D})  - \mu^2 +i\veps }\right] 
\right\}.
\nn
\eea
Furthermore, define
\begin{subequations}
\bea
{\cal I}_n & = &  \int\limits_0^1 dy \, y^n \, 
\ln[ (r/\mu)^2\,(1-y)^2 + y ], \\
{\cal J}_n & = &  \int\limits_0^1 dy \, y^n \, 
\ln [ 1 + (\bar Q/\mu)^2\, y(1-y)],  \\
{\cal K}_n & = &  \int\limits_0^1 dy \, y^n \, {\cal D} \, \ln  \frac{(1-y)(  \be - y\be_\ga - y \bar Q^2 + {\cal D})  - \mu^2 }{(1-y)
(  \be - y\be_\ga - y \bar Q^2 - {\cal D})  - \mu^2 }, \\
{\cal L}_n & = &   \int\limits_0^1 dy \, 
\frac{y^n}{{\cal D}} \, \ln  \frac{(1-y)(  \be - y\be_\ga - y \bar Q^2 + {\cal D})  - \mu^2 }{(1-y)
(  \be - y\be_\ga - y \bar Q^2 - {\cal D})  - \mu^2 }, 
\eea
\end{subequations}
Then,
\bea
g_M^{(e)} &=&  C_{N\De}\left\{ \third (1-\ln \mu) - \al  {\cal I}_1 + \al_\ga {\cal I}_2 
- \be  {\cal J}_1 + \be_\ga {\cal J}_2 -  {\cal K}_1 \right\}, \\
g_E^{(e)} &=& - C_{N\De}\left\{ - \third \ln \mu - (\be-r)  {\cal I}_1 + (\be_\ga-r) {\cal I}_2 
- (\al+r)  {\cal J}_1 + (\al_\ga+r) {\cal J}_2 \right. \nn\\
&+& \left. \be (\al+r) {\cal L}_1 - [ \be (\al_\ga+r) +  \be_\ga (\al+r) ]  
{\cal L}_2 + \be_\ga (\al_\ga+r) {\cal L}_3 \right\}, \\
g_C^{(e)} &=&  \half C_{N\De}\left\{ 2{\cal I}_2 - {\cal I}_1-2 {\cal J}_2 + {\cal J}_1  
+ (\al +r )  ( {\cal L}_1 - 2{\cal L}_2) - (\al_\ga +r )  ( {\cal L}_2 - 2{\cal L}_3)\right\},
\eea

\begin{subequations}
\bea
{\cal I}_1 & = & \frac{8 - 6(\mu/r)^2
+ (\mu/r)^4 }{2\sqrt{4 (r/\mu)^2 - 1}} \, \arctan \sqrt{4 (r/\mu)^2 - 1} \nn\\
&& - \thalf + \half (\mu/r)^2 
+ \left[ 1 - 2 (\mu/r)^2 + \half (\mu/r)^2 \right] \, \ln (r/\mu)  \,,\\
{\cal I}_2 & = & \frac{4 -\frac{19}{3} (\mu/r)^2 + \frac{8}{3}(\mu/r)^4- \third (\mu/r)^6 }{\sqrt{4 (r/\mu)^2 - 1}} 
\arctan \sqrt{4 (r/\mu)^2 - 1} \nn \\
&& - \mbox{$\frac{11}{9}$} + \mbox{$\frac{3}{2}$} (\mu/r)^2 - \third (\mu/r)^4 + \left[ \mbox{$\frac{2}{3}$}
 - 3 (\mu/r)^2 + 2 (\mu/r)^4 \right] \, \ln (r/\mu)\,,\\
{\cal J}_1 & = & -1 + \frac{\sqrt{4\mu^2 + \bar Q^2}}{2 \bar Q} \,
\ln \frac{   \sqrt{4\mu^2 + \bar Q^2}+ \bar Q}{ \sqrt{4\mu^2 + \bar Q^2}- \bar Q}\,,\\
{\cal J}_2 &=& -\frac{13}{18} - \frac{2 \mu^2}{3\bar Q^2} + \frac{2 ( \frac{\bar Q}{\mu} 
+ \frac{5 \mu}{\bar Q} +\frac{4 \mu^3}{\bar Q^3})}{3\sqrt{4\mu^2 + \bar Q^2}}\,
 \ln \frac{   \sqrt{4\mu^2 + \bar Q^2}+ \bar Q}{ \sqrt{4\mu^2 + \bar Q^2}- \bar Q }\,.
\eea
\end{subequations}

\section{Absorptive parts of the vertex corrections}
\label{sec:phase}

In the sideways dispersion relation method, we evaluate the 
$\pi N$ loop contribution to the $\gamma N \Delta$ form factors using 
a dispersion relation in the invariant mass of the $\Delta$. 
To present the results of this technique, we first 
introduce a number of kinematical quantities and notations.  
The pion momentum in the rest frame of the decaying $\Delta$ with invariant 
mass $W=\sqrt{s}$ is given by 
$|{\bf p}_\pi| \,=\, W \, \tilde \lambda$, 
with 
\begin{eqnarray}
\tilde \lambda &=& \half 
\sqrt{\left[1 - (\tilde r + \tilde \mu)^2 \right]\, 
\left[ 1 - (\tilde r - \tilde \mu)^2 \right]} \, ,
\nonumber \\
&=& \half 
\sqrt{\left[(1 + \tilde r)^2 - \tilde \mu^2 \right]\, 
\left[ (1 - \tilde r)^2 - \tilde \mu^2 \right]} \, ,
\label{eq:lambda}
\end{eqnarray}
and
$\tilde \mu = m_\pi/W$, $\tilde r = M_N/W$.
We also introduce 
$r_\Delta = M_\Delta/W$. 
In the on-shell case, $W = M_\Delta$, we of course have 
$\tilde r = r$, $\tilde \mu = \mu$, and $\tilde \lambda = \lambda$.

Furthermore, the energies of the pion ($E_\pi$) and nucleon ($E_N$) in the 
rest frame of the decaying $\Delta$ can be expressed as
$E_\pi = W \tilde \beta$ and $E_N = W \tilde \alpha$ with the fractions given
by:
\begin{eqnarray}
\tilde \beta = \half \, 
\left( 1 \,-\, \tilde r^2 + \tilde \mu^2 \right) , 
\hspace{2cm}
\tilde \alpha = \half \, 
\left( 1 \,+\, \tilde r^2 - \tilde \mu^2 \right) = 1 - \tilde \beta . 
\label{eq:pi3}
\end{eqnarray}
and the Lorentz-invariant pion (nucleon) velocities, denoted by  
$v_\pi$ ($v_N$) respectively are given by:
\begin{eqnarray}
v_\pi \equiv \frac{|{\bf p}_\pi|}{E_\pi} =
\frac{\tilde \lambda}{\tilde \beta} , 
\hspace{2cm}
v_\N \equiv \frac{|{\bf p}_N|}{E_N} = 
\frac{\tilde \lambda}{\tilde \alpha} ,
\end{eqnarray}
where we used $|{\bf p}_N| = |{\bf p}_\pi|$ in the $\Delta$ rest frame. 

We denote the virtual photon four-momentum by 
$q = (q^0, {\bf q})$. For spacelike virtual photons,  
the virtuality is usually denoted by $Q^2 \equiv {\bf q}^2 - q_0^2 > 0$.  
The virtual photon three-momentum 
in the $\Delta$ rest frame can then be expressed as 
$|{\bf q}| \,=\, W \, \tilde \kappa$, with 
\begin{eqnarray}
\tilde \kappa \,&=&\, \half
\sqrt{(1 - \tilde r^2)^2 + 2 \tilde Q^2 (1 + \tilde r^2) + \tilde Q^4 } \, ,
\nonumber \\
&=&\, \half
\sqrt{\left[(1 + \tilde r)^2 + \tilde Q^2 \right]\, 
\left[ (1 - \tilde r)^2 + \tilde Q^2 \right]} \, ,
\label{eq:kappa}
\end{eqnarray}
where 
$\tilde Q^2 \equiv Q^2/s$.
It is also useful to introduce the fraction of the photon energy:  
\begin{eqnarray}
\tilde \beta_\gamma \equiv \frac{p^\prime \cdot q}{s}  = 
\half \left( 1 - \tilde r^2 - \tilde Q^2 \right),
\end{eqnarray} 
satisfying $\kappa^2 = \tilde \beta_\gamma^2 + \tilde Q^2$. 
Finally, we define 
\begin{eqnarray}
w_\pi &\equiv& 
\frac{2 \,|{\bf p}_\pi| \,|{\bf q}|  }{Q^2 + 2 \, E_\pi \, q^0} 
\,=\, \frac{2 \, \tilde \kappa \, \tilde \lambda}{\tilde Q^2 
+ 2 \, \tilde \beta \, \tilde \beta_\gamma}, \\
w_N &\equiv& 
\frac{2 \,|{\bf p}_N| \,|{\bf q}|  }{Q^2 + 2 \, E_N \, q^0} 
\,=\, \frac{2 \, \tilde \kappa \, \tilde \lambda}{\tilde Q^2 
+ 2 \, \tilde \alpha \, \tilde \beta_\gamma}. 
\end{eqnarray}
In the real photon limit ($Q^2 \to 0$), one obtains that $w_\pi \to v_\pi$, 
and $w_N \to v_N$.

The calculation of the absorptive part of these 
$\pi N$ loop contribution, entering the sideways dispersion relations, 
involves several phase space integrals, of which the simplest is of the form~:
\begin{eqnarray}
\int \frac{d^4 p_\pi}{(2 \pi)^4} \, 
\left[ (2 \pi) \, \delta(p_\pi^2 - m_\pi^2) \, \Theta(p_\pi^0) \right] \,
\left[ (2 \pi) \, \delta((p^\prime - p_\pi)^2 - M_N^2) \, 
\Theta(p^{\prime \, 0} - p_\pi^0) \right] 
\;=\; \frac{\tilde \lambda}{4 \, \pi}.
\end{eqnarray}
\newline
\indent
The evaluation of the absorptive part corresponding with 
Fig.~\ref{fig:gandelabs} (a), firstly 
involves the following phase space integrals with different powers of 
pion momentum in the numerator~:
\begin{eqnarray}
&&\int \frac{d^4 p_\pi}{(2 \pi)^4} \, 
\left[ (2 \pi) \, \delta(p_\pi^2 - m_\pi^2) \, \Theta(p_\pi^0) \right] \,
\left[ (2 \pi) \, \delta((p^\prime - p_\pi)^2 - M_N^2) \, 
\Theta(p^{\prime \, 0} - p_\pi^0) \right] \,
p_\pi^\alpha 
\nonumber \\
&&\;=\;
a_{1 \, ,\pi} \, p^{\prime \, \alpha} , \\
\nonumber \\
&&\int \frac{d^4 p_\pi}{(2 \pi)^4} \, 
\left[ (2 \pi) \, \delta(p_\pi^2 - m_\pi^2) \, \Theta(p_\pi^0) \right] \,
\left[ (2 \pi) \, \delta((p^\prime - p_\pi)^2 - M_N^2) \, 
\Theta(p^{\prime \, 0} - p_\pi^0) \right] \,
p_\pi^\alpha \, p_\pi^\beta
\nonumber \\
&&\;=\;
a_{2 \, ,\pi} \, p^{\prime \, 2} \, g^{\alpha \beta} \,+\, 
a_{3 \, ,\pi} \, p^{\prime \, \alpha} \, p^{\prime \, \beta} , 
\end{eqnarray}
with
\begin{eqnarray}
a_{1 \, ,\pi} &\,=\,& \frac{\tilde \lambda}{4 \, \pi} \, 
\tilde \beta,\\
a_{2 \, ,\pi} &\,=\,& \frac{1}{4 \, \pi} \, 
\left(- \frac{\tilde \lambda^3}{3} \right) ,\\
a_{3 \, ,\pi} &\,=\,& \frac{1}{4 \, \pi} \, 
\tilde \lambda \, \left( \tilde \beta^2 + \frac{\tilde \lambda^2}{3} \right) .
\end{eqnarray}

Furthermore, the calculation of the absorptive part corresponding with 
Fig.~\ref{fig:gandelabs} (a), which has one 
pion which is off-shell, also involves the following integrals~:
\begin{eqnarray}
&&\int \frac{d^4 p_\pi}{(2 \pi)^4} \, 
\left[ (2 \pi) \, \delta(p_\pi^2 - m_\pi^2) \, \Theta(p_\pi^0) \right] \,
\left[ (2 \pi) \, \delta((p^\prime - p_\pi)^2 - M_N^2) \, 
\Theta(p^{\prime \, 0} - p_\pi^0) \right] \,
\frac{1}{q^2 - 2 \, p_\pi \cdot q}
\nonumber \\
&&\;=\;
- \frac{1}{2 \, p^{\prime \, 2}} \, b_{1 \, ,\pi} , 
\label{eq:abspi0} 
\end{eqnarray}
\begin{eqnarray}
&&\int \frac{d^4 p_\pi}{(2 \pi)^4} \, 
\left[ (2 \pi) \, \delta(p_\pi^2 - m_\pi^2) \, \Theta(p_\pi^0) \right] \,
\left[ (2 \pi) \, \delta((p^\prime - p_\pi)^2 - M_N^2) \, 
\Theta(p^{\prime \, 0} - p_\pi^0) \right] \,
\frac{p_\pi^\alpha}{q^2 - 2 \, p_\pi \cdot q}
\nonumber \\
&&\;=\; - \frac{1}{2 \, p^{\prime \, 2}} \left\{ 
b_{2 \, ,\pi} \, p^{\prime \, \alpha} \,+\, 
b_{3 \, ,\pi} \, q^\alpha \right\} , 
\label{eq:abspi1} 
\end{eqnarray}
\begin{eqnarray}
&&\int \frac{d^4 p_\pi}{(2 \pi)^4} \, 
\left[ (2 \pi) \, \delta(p_\pi^2 - m_\pi^2) \, \Theta(p_\pi^0) \right] \,
\left[ (2 \pi) \, \delta((p^\prime - p_\pi)^2 - M_N^2) \, 
\Theta(p^{\prime \, 0} - p_\pi^0) \right] \,
\frac{p_\pi^\alpha \, p_\pi^\beta}{q^2 - 2 \, p_\pi \cdot q}
\nonumber \\
&&\;=\; - \frac{1}{2} \left\{ 
c_{1 \, , \pi} \, g^{\alpha \beta} 
\,+\, c^\prime_{1 \, , \pi} \, 
\left( \frac{q^\alpha p^{\prime \, \beta} + q^\beta p^{\prime \, \alpha}}
{q \cdot p^\prime} \right)
\,+\, \frac{c_{2 \, ,\pi}}{p^{\prime \, 2}} 
\, p^{\prime \, \alpha} p^{\prime \, \beta} 
\,+\, \frac{c_{3 \, ,\pi}}{p^{\prime \, 2}} \, q^\alpha q^\beta  
\right\}, 
\label{eq:abspi2} 
\end{eqnarray}
\begin{eqnarray}
&&\int \frac{d^4 p_\pi}{(2 \pi)^4} \, 
\left[ (2 \pi) \, \delta(p_\pi^2 - m_\pi^2) \, \Theta(p_\pi^0) \right] \,
\left[ (2 \pi) \, \delta((p^\prime - p_\pi)^2 - M_N^2) \, 
\Theta(p^{\prime \, 0} - p_\pi^0) \right] \,
\frac{p_\pi^\alpha \, p_\pi^\beta \, p_\pi^\gamma}{q^2 - 2 \, p_\pi \cdot q}
\nonumber \\
&&\;=\;- \frac{1}{2} \left\{\frac{}{} 
d_{1 \, , \pi} \, \left[ g^{\alpha \beta} \, p^{\prime \, \gamma} \,+\,  
g^{\beta \gamma} \, p^{\prime \, \alpha} \,+\,  
g^{\gamma \alpha} \, p^{\prime \, \beta} \right]  \right. \nonumber \\
&&\hspace{1.25cm} \,+\, 
d^\prime_{1 \, , \pi} \, \frac{2}{q \cdot p^\prime}
\left[  q^\alpha  p^{\prime \, \beta} p^{\prime \, \gamma} 
+ q^\beta  p^{\prime \, \alpha}  p^{\prime \, \gamma} \,+\,  
+ q^\gamma p^{\prime \, \alpha} p^{\prime \, \beta} 
\right] 
\nonumber \\
&&\hspace{1.25cm} \,+\, 
d_{2 \, , \pi} \, \left( g^{\alpha \beta} \, q^\gamma \,+\,  
g^{\beta \gamma} \, q^\alpha \,+\,  
g^{\gamma \alpha} \, q^\beta \right)  \nonumber \\
&&\hspace{1.25cm} \,+\, 
d^\prime_{2 \, , \pi} \, \frac{2}{q \cdot p^\prime}
\left[  q^\alpha q^\beta p^{\prime \, \gamma} 
+ q^\beta q^\gamma p^{\prime \, \alpha} 
+ q^\gamma q^\alpha p^{\prime \, \beta} \right]
\nonumber \\
&&\hspace{1.25cm} \left. \,+\, 
\frac{d_{3 \, , \pi}}{p^{\prime \, 2}} 
\, p^{\prime \, \alpha} p^{\prime \, \beta} p^{\prime \, \gamma} 
\,+\, \frac{d_{4 \, , \pi}}{p^{\prime \, 2}} \, q^\alpha q^\beta q^\gamma 
\right\} .
\label{eq:abspi3}  
\end{eqnarray}
\newline
\indent
The coefficients appearing in the above equations are given by~:
\begin{eqnarray}
b_{1 \, ,\pi} &=& \frac{1}{4 \pi}  
\frac{1}{2 \tilde \kappa} \, 
\log \left( \frac{1 + w_\pi}{1 - w_\pi} \right)  ,
\label{eq:b1}
\\
b_{2 \, ,\pi} &=& \frac{1}{4 \pi} \frac{1}{4 \tilde \kappa^2}    
\left\{ 4 \, \tilde \lambda \, \tilde \beta_\gamma 
+ \tilde Q^2 
\left( 2 \tilde \beta - \tilde \beta_\gamma \right)  \frac{1}{\tilde \kappa} \,
\log \left( \frac{1 + w_\pi}{1 - w_\pi} \right) \right\} , 
\\
b_{3 \, ,\pi} &=& \frac{1}{4 \pi}  \frac{\tilde \lambda}{2 \tilde \kappa^2}   
\left\{ - 2
+ \frac{1}{w_\pi} \,\log \left( \frac{1 + w_\pi}{1 - w_\pi} \right) \right\} , 
\end{eqnarray}
\begin{eqnarray}
c_{1 \, ,\pi} &=& \frac{1}{4  \pi} 
\frac{\tilde \lambda^2}{4  \tilde \kappa}   
\left\{ - \frac{2}{w_\pi} 
+ \log \left( \frac{1 + w_\pi}{1 - w_\pi} \right) 
\left[ -1 + \frac{1}{w_\pi^2} \right] \right\} ,
\\
c^\prime_{1 \, ,\pi} &=& \frac{1}{4 \pi} 
\frac{1}{4 \tilde \kappa^2} 
\left\{ \tilde \lambda \left[ 
\left( 1 - \frac{3 \, \tilde Q^2}{\tilde \kappa^2} \right) 
\left( 2 \, \tilde \beta \tilde \beta_\gamma  + \tilde Q^2 \right) 
+ 2 \, \tilde Q^2 \right]  \right. \nonumber \\ 
&+&\left. \frac{1}{2  \tilde \kappa} 
\log \left( \frac{1 + w_\pi}{1 - w_\pi} \right)  
\left[ \tilde \lambda^2 \left( 2 \tilde \beta_\gamma^2 
\left(1 - \frac{1}{w_\pi^2} \right) + \frac{4 \tilde Q^2}{w_\pi^2} \right) 
- \tilde Q^2 (2  \tilde \beta  \tilde \beta_\gamma + \tilde Q^2)
\right]  \right\} ,
\\
c_{2 \, ,\pi} &=& \frac{1}{4 \pi}  \frac{1}{4 \tilde \kappa^2}   
\left\{ \tilde \lambda \left[ \left( 1 + \frac{3 \, \tilde Q^2}{2\, \tilde \kappa^2} \right) 2 
\left( 2 \tilde \beta \tilde \beta_\gamma + \tilde Q^2 \right) 
- 4 \, \tilde Q^2 \right]  
\right. \nonumber \\ 
&+&\left. \frac{\tilde Q^2}{2 \tilde \kappa} 
\log \left( \frac{1 + w_\pi}{1 - w_\pi} \right) 
\left[ 2 \tilde \lambda^2 \left( 1 - \frac{3}{w_\pi^2} \right)
+ 4 \tilde \beta^2 + \tilde Q^2 
\right] \right\} ,
\\
c_{3 \, ,\pi} &=& \frac{1}{4 \pi}  
\frac{\tilde \lambda^2}{4 \tilde \kappa^3}   
\left\{ - \frac{6}{w_\pi} 
+ \log \left( \frac{1 + w_\pi}{1 - w_\pi} \right) 
\left[ - 1 + \frac{3}{w_\pi^2} \right] \right\}, 
\end{eqnarray}
\begin{eqnarray}
d_{1 \, ,\pi} &=& \frac{1}{4 \pi} \frac{\tilde \lambda^2}{4 \tilde \kappa^3}   
\left\{   
- \frac{4}{3} \tilde \kappa \tilde \lambda \tilde \beta_\gamma 
- \frac{\tilde Q^2 (2 \tilde \beta - \tilde \beta_\gamma)}{w_\pi} 
+ \frac{\tilde Q^2}{2} 
\log \left( \frac{1 + w_\pi}{1 - w_\pi} \right)  
(2 \tilde \beta - \tilde \beta_\gamma ) 
\left[- 1 + \frac{1}{w_\pi^2} \right] \right\}, \;\;\;\;\;\;\;\;\;  
\\
d^\prime_{1 \, ,\pi} &=& \frac{1}{4 \pi} 
\frac{\tilde \beta_\gamma}{8 \tilde \kappa^4}
\left\{ \tilde \lambda \left[  2 \tilde \lambda^2 
\left( \frac{2}{3} (\tilde \beta_\gamma^2 - \tilde Q^2)  
+ \frac{5 \tilde Q^2}{w_\pi^2} \right) 
- \tilde Q^2 \left( 4 \tilde \beta^2 + 2 \tilde Q^2 
+ 2 \tilde \beta \tilde \beta_\gamma \right) \right] \right.
\nonumber \\
&+&  \frac{\tilde Q^2}{2 \tilde \kappa} 
\log \left( \frac{1 + w_\pi}{1 - w_\pi} \right)  
\left[\tilde \lambda^2 \left( 
\tilde \beta_\gamma (2 \tilde \beta - \tilde \beta_\gamma ) 
\left(3 - \frac{5}{w_\pi^2} \right)
+ \tilde \kappa^2 \left(1 - \frac{1}{w_\pi^2} \right) \right) 
\right. \nonumber \\
&&\left.\left. \hspace{3cm} 
+ \tilde \beta (2 \tilde \beta - \tilde \beta_\gamma )
(2 \tilde \beta \tilde \beta_\gamma + \tilde Q^2)
\right] \right\} ,  
\end{eqnarray}
\begin{eqnarray}
d_{2 \, ,\pi} &=& \frac{1}{4 \pi} \frac{\tilde \lambda^3}{2 \tilde \kappa^2}   
\left\{ \frac{2}{3}   - \frac{1}{w_\pi^2}  
+  \frac{1}{2 w_\pi} 
\log \left( \frac{1 + w_\pi}{1 - w_\pi} \right)   
\left[- 1 + \frac{1}{w_\pi^2} \right] \right\} ,  
\\
d^\prime_{2 \, ,\pi} &=& \frac{1}{4 \pi} \frac{1}{8 \tilde \kappa^4} 
\left\{ \tilde \lambda \left[ 2 \tilde \lambda^2 
\left( \tilde \beta_\gamma^2 \left( -\frac{4}{3} + \frac{2}{w_\pi^2} \right) 
- \frac{3 \tilde Q^2}{w_\pi^2} \right) 
+ \frac{3}{2} \tilde Q^2 
\left( 2 \, \tilde \beta \tilde \beta_\gamma  + \tilde Q^2 \right)
\right]  \right.  \nonumber \\ 
&+& \left.  \frac{\tilde \lambda^2}{2  \tilde \kappa} 
\log \left( \frac{1 + w_\pi}{1 - w_\pi} \right)  
\left[ 4 \tilde \kappa^2 \tilde \beta \tilde \beta_\gamma
\left( 1 - \frac{1}{w_\pi^2} \right)
- \tilde \beta_\gamma \tilde Q^2 ( 2 \tilde \beta - \tilde \beta_\gamma ) 
\left(3 - \frac{5}{w_\pi^2} \right)
\right]  \right\} ,
\\
d_{3 \, ,\pi} &=& -3 \, d_{1 \, ,\pi} - 4 \, d^\prime_{1 \, ,\pi} 
- 2 \tilde \beta_\gamma ( d_{2 \, ,\pi} + d^\prime_{2 \, ,\pi} )
+ \tilde \beta^2 \, b_{2 \, , \pi} ,
\\
d_{4 \, ,\pi} &=& \frac{2}{\tilde Q^2} ( d_{2 \, ,\pi} + d^\prime_{2 \, ,\pi} )
+ \frac{1}{2} \, c_{3 \, , \pi} .
\label{eq:d4}
\end{eqnarray}
\newline
\indent
The calculation of the absorptive part corresponding with 
Fig.~\ref{fig:gandelabs} (b), which has one 
nucleon which is off-shell, involves the integrals ~:
\begin{eqnarray}
&&\int \frac{d^4 p_N}{(2 \pi)^4} \, 
\left[ (2 \pi) \, \delta(p_N^2 - M_N^2) \, \Theta(p_N^0) \right] \,
\left[ (2 \pi) \, \delta((p^\prime - p_N)^2 - m_\pi^2) \, 
\Theta(p^{\prime \, 0} - p_N^0) \right] \,
p_N^\alpha 
\nonumber \\
&&\;=\;
a_{1 \, ,N} \, p^{\prime \, \alpha} , \\
\nonumber \\
&&\int \frac{d^4 p_N}{(2 \pi)^4} \, 
\left[ (2 \pi) \, \delta(p_N^2 - M_N^2) \, \Theta(p_N^0) \right] \,
\left[ (2 \pi) \, \delta((p^\prime - p_N)^2 - m_\pi^2) \, 
\Theta(p^{\prime \, 0} - p_N^0) \right] \,
p_N^\alpha \, p_N^\beta
\nonumber \\
&&\;=\;
a_{2 \, ,N} \, p^{\prime \, 2} \, g^{\alpha \beta} \,+\, 
a_{3 \, ,N} \, p^{\prime \, \alpha} \, p^{\prime \, \beta} , 
\end{eqnarray}
\begin{eqnarray}
&&\int \frac{d^4 p_N}{(2 \pi)^4} \, 
\left[ (2 \pi) \, \delta(p_N^2 - M_N^2) \, \Theta(p_N^0) \right] \,
\left[ (2 \pi) \, \delta((p^\prime - p_N)^2 - m_\pi^2) \, 
\Theta(p^{\prime \, 0} - p_N^0) \right] \,
\frac{1}{q^2 - 2 \, p_N \cdot q}
\nonumber \\
&&\;=\;
- \frac{1}{2 \, p^{\prime \, 2}} \, b_{1 \, ,N} , 
\end{eqnarray}
and analogous integrals as in Eqs.~(\ref{eq:abspi1}-\ref{eq:abspi3}), where 
$p_\pi \leftrightarrow p_N$. 
\newline
\indent
The coefficients appearing in the above integrals are given by~:
\begin{eqnarray}
a_{1 \, ,N} &\,=\,& \frac{\tilde \lambda}{4 \, \pi} \, \tilde \alpha ,\\
a_{2 \, ,N} &\,=\,& \frac{1}{4 \, \pi} \, 
\left(- \frac{\tilde \lambda^3}{3} \right) ,\\
a_{3 \, ,N} &\,=\,& \frac{1}{4 \, \pi} \, 
\tilde \lambda \, \left( \tilde r^2 + \frac{4 \, \tilde \lambda^2}{3} \right) .
\end{eqnarray}
and
\begin{eqnarray}
b_{1 \, ,N} &=& \frac{1}{4 \pi} \, 
\frac{1}{2 \, \tilde \kappa} \, 
\log \left( \frac{1 + w_N}{1 - w_N} \right)  ,
\label{eq:b1n}\\
b_{2 \, ,N} &=& \frac{1}{4 \pi} \frac{1}{4 \, \tilde \kappa^2}    
\left\{ 4 \, \tilde \lambda \, \tilde \beta_\gamma 
+ \, \tilde Q^2 
\left[ 1 + \tilde r^2 + \tilde Q^2 + 2 (\tilde r^2 - \tilde \mu^2) \right]  
\frac{1}{2 \, \tilde \kappa} \,
\log \left( \frac{1 + w_N}{1 - w_N} \right) \right\} , \\
b_{3 \, , N} &=& \frac{1}{4 \pi}  \frac{1}{2 \, \tilde \kappa^2}   
\left\{ - 2 \, \tilde \lambda  
+ \left( 2 \tilde \alpha \tilde \beta_\gamma + \tilde Q^2 \right) 
\frac{1}{2 \, \tilde \kappa} \,
\log \left( \frac{1 + w_N}{1 - w_N} \right) \right\} , 
\end{eqnarray}
\begin{eqnarray}
c_{1 \, ,N} &=& \frac{1}{4 \tilde \kappa^2}   
\left\{- 2 \tilde \beta_\gamma  a_{1 \, , N}  
+ 2 \left( \tilde \beta_\gamma^2 \, \tilde r^2 
- \tilde Q^2  \tilde \lambda^2 \right)  b_{1 \, , N} 
+ \tilde Q^2  \tilde \beta_\gamma  b_{2 \, , N} 
+ \tilde Q^2  \left(\tilde \kappa^2 + \tilde \beta_\gamma^2 \right) 
\, b_{3 \, , N} \right\} , \;\;\; \\
c^\prime_{1 \, ,N} &=& - \left( 1 - \frac{3 \, \tilde Q^2}{\tilde \kappa^2}\right) 
\, c_{1 \, , N} + \tilde Q^2 \frac{\tilde \lambda^2}{\tilde \kappa^2} \, b_{1 \, , N} 
- \frac{\tilde Q^2}{2} \, b_{3 \, , N} , \\
c_{2 \, ,N} &=& - 2 \, c_{1 \, , N} - \tilde Q^2 \, c_{3 \, , N} 
+ \tilde r^2 \, b_{1 \, , N} + \tilde Q^2 \, b_{3 \, , N} , \\
c_{3 \, ,N} &=& \frac{1}{\tilde \kappa^2} \, 
\left\{ 3 \, c_{1 \, , N} + \tilde \lambda^2 \, b_{1 \, , N} \right\} , 
\end{eqnarray}
\begin{eqnarray}
d_{1 \, ,N} &=& \frac{1}{2 \, \tilde \kappa^2}   
\left\{- \tilde \beta_\gamma \, (2 \, a_{2 \, , N} + a_{3 \, , N})  
+ \left[ \tilde r^2 \, (\tilde \kappa^2 - \tilde Q^2)  
- \tilde Q^2 \, \tilde \lambda^2 \right] \, b_{2 \, , N} \right. \nonumber \\
&+&\left. \frac{\tilde Q^2}{2 \tilde \beta_\gamma} \, 
\left[ 2 \, \tilde \kappa^2 - \tilde Q^2 \right] \, c^\prime_{1 \, , N} 
+ \frac{\tilde Q^2}{2} \, \tilde \beta_\gamma  
\, (c_{2 \, , N} + 2 \, c_{1 \, ,N} ) \right\} ,\\
d^\prime_{1 \, ,N} &=& - \frac{5}{2} \, d_{1 \, ,N} 
+ \frac{\tilde r^2}{2} \, b_{2 \, ,N} 
- \frac{(1 - \tilde r^2)}{4 \, \tilde \kappa^2} \, a_{3 \, , N} \nonumber \\
&+& \frac{\tilde Q^2}{2 \, \tilde \kappa^2} 
\left[ 5 \, d_{1 \, ,N} + \frac{1}{2} \, a_{3\, , N} 
- \tilde r^2 \, b_{2 \, , N} + \frac{1}{2} \tilde \beta_\gamma 
\,( c^\prime_{1 \, , N} + c_{2 \, ,N}) \right] , \\
d_{2 \, ,N} &=& \frac{1}{\tilde Q^2}\, \tilde \beta_\gamma \, 
d_{1 \, , N} + \frac{1}{2} \, c_{1 \, , N} 
- \frac{1}{\tilde Q^2} \, a_{2 \, , N} , \\
d^\prime_{2 \, ,N} &=& \frac{1}{2 \, \tilde Q^2}\, \tilde \beta_\gamma \, 
(d_{1 \, , N} + 2 \, d^\prime_{1 \, , N}) + \frac{1}{4} \, c^\prime_{1 \, , N},
\\
d_{3 \, ,N} &=& - 3 \, d_{1 \, ,N} - 4 \, d^\prime_{1 \, ,N}
- 2 \tilde \beta_\gamma \, (d_{2 \, , N} + d^\prime_{2 \, , N}) 
+ \tilde \alpha^2 \, b_{2 \, , N}, \\
d_{4 \, ,N} &=& \frac{2}{\tilde Q^2} \, (d_{2 \, , N} + d^\prime_{2 \, , N}) 
+ \frac{1}{2} \, c_{3 \, , N} . 
\label{eq:d4n}
\end{eqnarray}

The imaginary parts of $g_1$, $g_2$, and $g_3$ 
corresponding with the diagram of Fig.~\ref{fig:gandelabs} (a) are 
thus given by:
\begin{eqnarray}
\eqlab{gipi}
{\rm Im} \, g^+_{1,\, a} &=&  \,  
\,\frac{g_{A}\, h_{A} \, M_N}{4\, f_\pi^2} \, 
\, \left\{ \, d_{2 \, , \pi} \, \right\} 
\, , \nn \\
{\rm Im} \, g^+_{2, \, a} &=& -  \, 
\,\frac{g_{A}\, h_{A} \, M_N}{4 \, f_\pi^2} \, 
 \frac{1}{W}  
\left\{ \frac{\left( r_\Delta + \tilde r \right)}
{\tilde \beta_\gamma} \, 
( 2 \, d^\prime_{2 \, , \pi} - c^\prime_{1 \, , \pi}) 
\,+\, \frac{2 \, r_\Delta}{\tilde \beta_\gamma} 
\, d^\prime_{1 \, , \pi} \right\} , \\
{\rm Im} \, g^+_{3, \, a} &=&   
\,\frac{g_{A}\, h_{A} \, M_N}{4 \, f_\pi^2} \, 
 \frac{1}{W} 
\left\{ \frac{\left( r_\Delta + \tilde r \right)}{2} \, 
( b_{3 \, , \pi} - 3 \, c_{3 \, , \pi} + 2 \, d_{4 \, , \pi}) 
\,+\, \frac{r_\Delta}{2 \, \tilde \beta_\gamma} \, 
(4 \, d^\prime_{2 \, , \pi} - c^\prime_{1 \, , \pi})
\right\} . \nn
\end{eqnarray}

The imaginary parts of $g_1$, $g_2$, and $g_3$ 
corresponding with the diagram of Fig.~\ref{fig:gandelabs} (b) are 
likewise given by:
\begin{eqnarray}
\eqlab{giN}
{\rm Im} \, g^+_{1,\, b} &=& -   
\,\frac{g_{A}\, h_{A} \, M_N}{4 \, f_\pi^2} \, 
\left\{  d_{2 \, N}
+ \frac{1}{2} \tilde r (r_\Delta - \tilde r) \, b_{3 \, , N} 
+ \frac{(1 - \tilde r \, r_\Delta)}{2 \tilde \beta_\gamma} \, 
c^\prime_{1 \, ,N} 
- \frac{\tilde Q^2}{2} \, c_{3 \, , N} \right\} 
\, ,\nn \\
{\rm Im} \, g^+_{2, \, b} &=&   
\,\frac{g_{A}\, h_{A} \, M_N}{4 \, f_\pi^2 \,W} \, 
\left\{  
- \tilde r \, b_{3 \, , N} 
- \frac{(r_\Delta - \tilde r)}{\tilde \beta_\gamma} 
\, c^\prime_{1 \, ,N} 
+ \frac{2 \, r_\Delta}{\tilde \beta_\gamma} \, d^\prime_{1 \, , N} 
+ \frac{2 \, (r_\Delta + \tilde r)}{\tilde \beta_\gamma} \, 
d^\prime_{2 \, , N} 
\right\}, \\
{\rm Im} \, g^+_{3, \, b} &=&   
\,\frac{g_{A}\, h_{A} \, M_N}{4 \, f_\pi^2} \, 
\frac{1}{W} \, \left\{  
 r_\Delta \, c_{3 \, , N} 
- \frac{2 \, r_\Delta}{\tilde \beta_\gamma} \, d^\prime_{2 \, , N} 
- (r_\Delta + \tilde r) \, d_{4 \, , N} 
\right\} \nn
\, . 
\end{eqnarray}

\begin{acknowledgments}
We thank C.~Alexandrou for helpful discussions of
the lattice QCD results, and 
C.-W. Kao for drawing our attention to Ref.~\cite{Kao:1999ig}.
This work is supported in part by DOE grant no.\
DE-FG02-04ER41302 and contract DE-AC05-84ER-40150 under
which SURA
operates the Jefferson Laboratory.  
\end{acknowledgments}

\newpage

\begin{figure}
\centerline{  \epsfxsize=3cm
  \epsffile{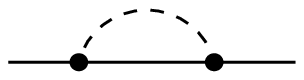} 
}
\caption{Nucleon self-energy to one loop.}
\figlab{nucself}
\end{figure}

\begin{figure}
\centerline{  \epsfxsize=11cm
  \epsffile{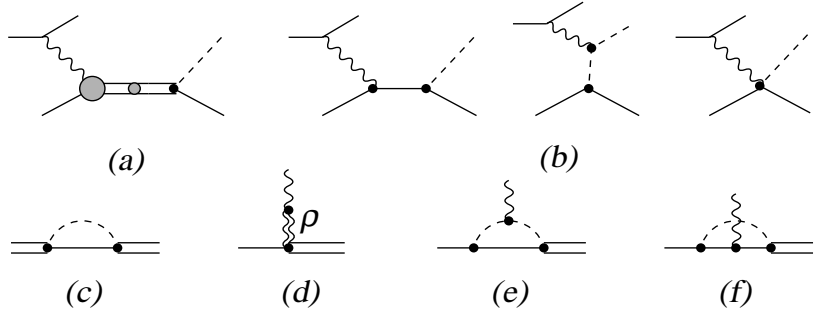} 
}
\caption{Diagrams for the $e N \to e \pi N $ reaction 
at NLO in the $\delta$-expansion, considered in this work. Double lines represent
the $\De$ propagators.}
\figlab{diagrams}
\end{figure}

\begin{figure}[h,b,t,p]
\centerline{
\epsfxsize=4.5cm
\epsffile{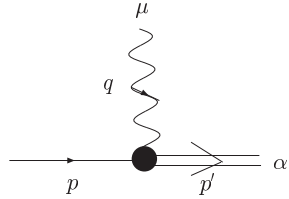}
}
\caption{$\gamma^* N \Delta$ vertex. The four-momenta of 
the nucleon ($\Delta$) and of the photon are given by 
$p$ ($p^\prime$) and $q \equiv p^\prime - p$ respectively. 
The four-vector index of 
the spin 3/2 field is given by $\alpha$, and 
$\mu$ is the four-vector index of the photon field.}
\figlab{treevertex}
\end{figure}

\begin{figure}
\centerline{  \epsfxsize=4cm
  \epsffile{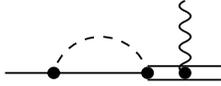} 
}
\caption{The higher-order $\ga N\De$ vertex correction
required by the electromagnetic current conservation.}
\figlab{extra}
\end{figure}

\begin{figure}[t,p]
\centerline{
\epsfxsize=10cm
\epsffile{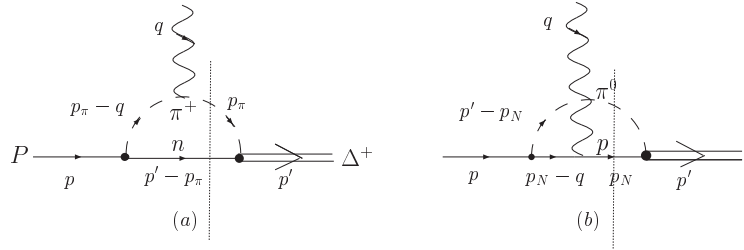}
}
\caption{Absorptive part of the $\gamma p \Delta^+$ vertex. 
Diagram (a) : $\pi^+ n$ loop where the photon couples to the $\pi^+$; 
diagram (b) : $\pi^0 p$ loop where the photon couples to the charge of the 
proton.  
The vertical dotted lines indicate that the $\pi N$ 
intermediate state is taken on-shell.}
\figlab{gandelabs}
\end{figure}

\begin{figure}[t,h]
\centerline{  \epsfxsize=13cm
  \epsffile{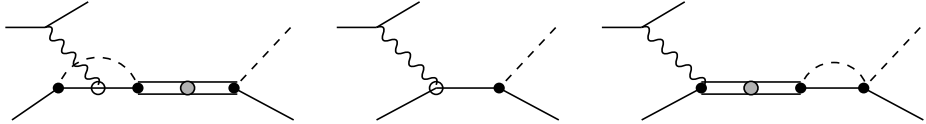} 
}
\caption{Examples of N$^2$LO contributions to
the $e N \to e \pi N $ reaction neglected in this work.
Open circles denote the \EM\ coupling to the
anomalous magnetic moment of the nucleon. }
\figlab{NNLOgraphs}
\end{figure}

\begin{figure}[t,h]
\includegraphics[width=0.47\columnwidth]{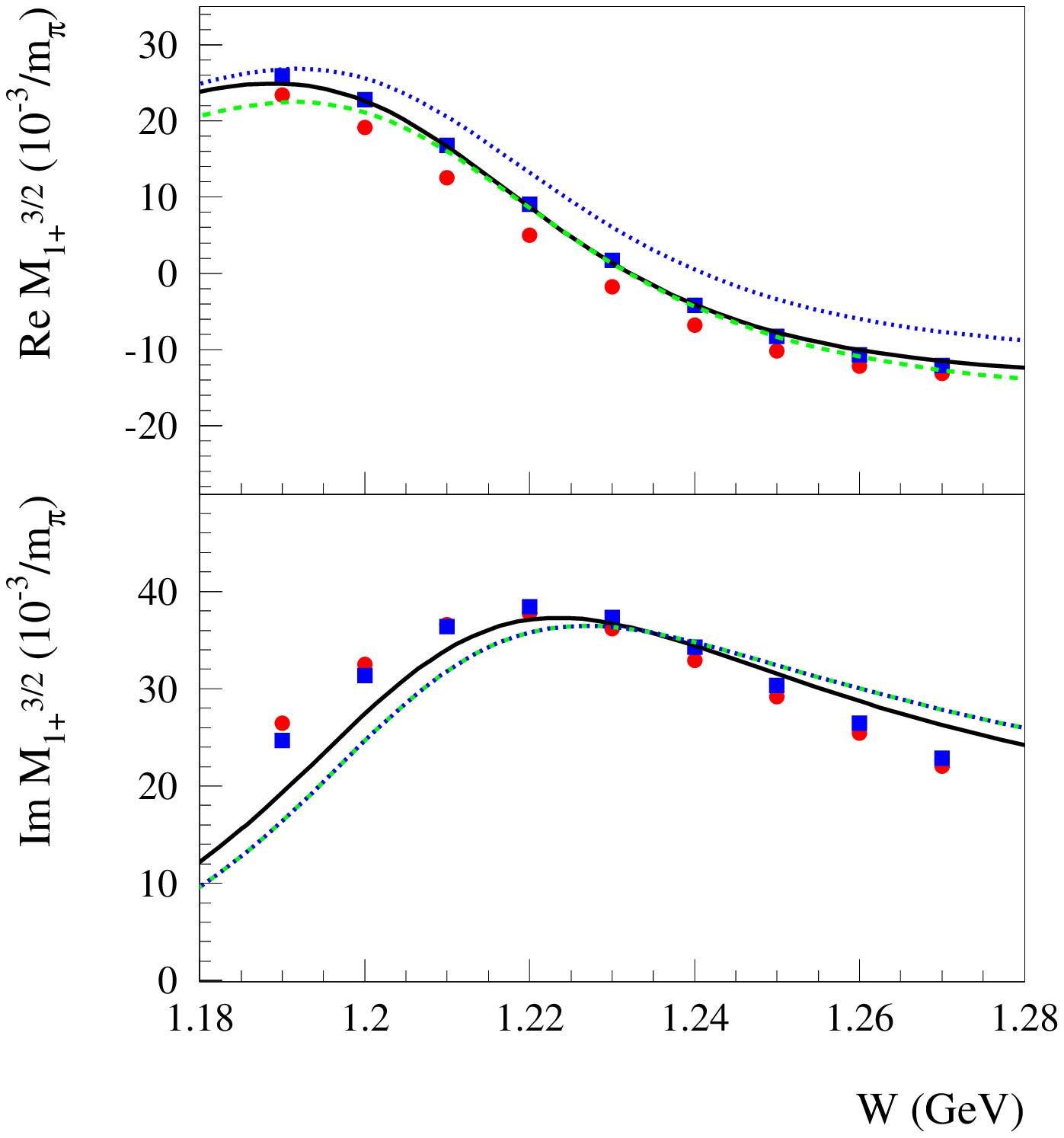}
\includegraphics[width=0.47\columnwidth]{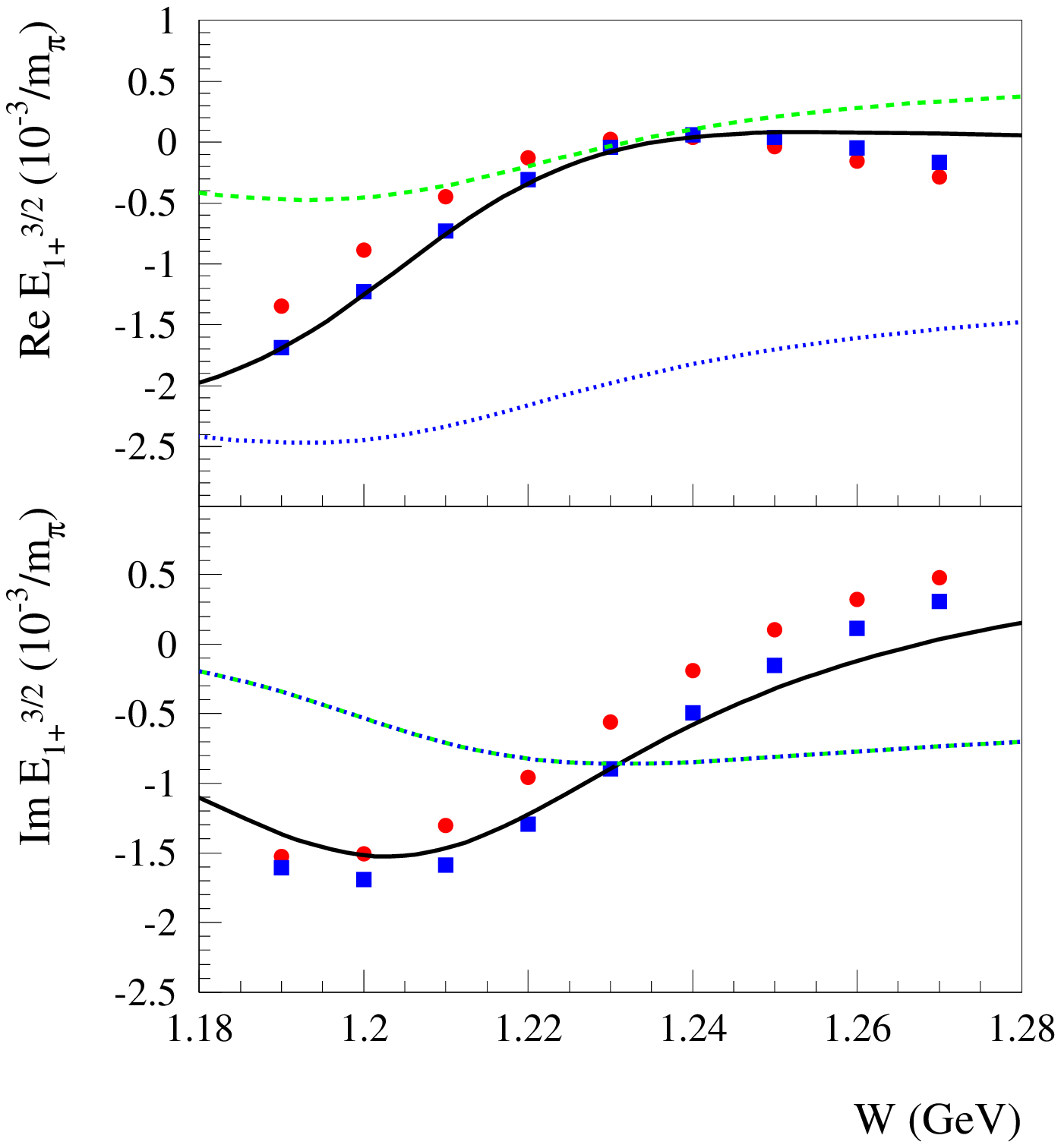}
\caption{
(Color online) Multipole amplitudes
$M_{1+}^{(3/2)}$ (left panel) and $E_{1+}^{(3/2)}$ (right panel)
for pion photoproduction as function of the invariant mass $W$ of 
the $\pi N$ system.  
Green dashed curves: 
$\Delta$ contribution without the $\ga N\De$-vertex corrections,
[i.e., \Figref{diagrams}(a) without  \Figref{diagrams}(e, f)].  
Blue dotted curves: adding the Born contributions, \Figref{diagrams}(b),
 to the dashed curves. 
Black solid curves: complete NLO calculation, includes 
all graphs from \Figref{diagrams}. In all curves
the low-energy parameters are chosen as~: 
$g_M = 2.9$, $g_E = -1.0$. 
The data point are from the 
SAID analysis~(FA04K)~\protect\cite{Arndt:2002xv} (red circles), and from the 
MAID 2003 analysis~\protect\cite{Drechsel:1998hk} (blue squares).
}
\figlab{gap_pin_mult}
\end{figure}

\begin{figure}[t,h]
\includegraphics[width=0.7\columnwidth]{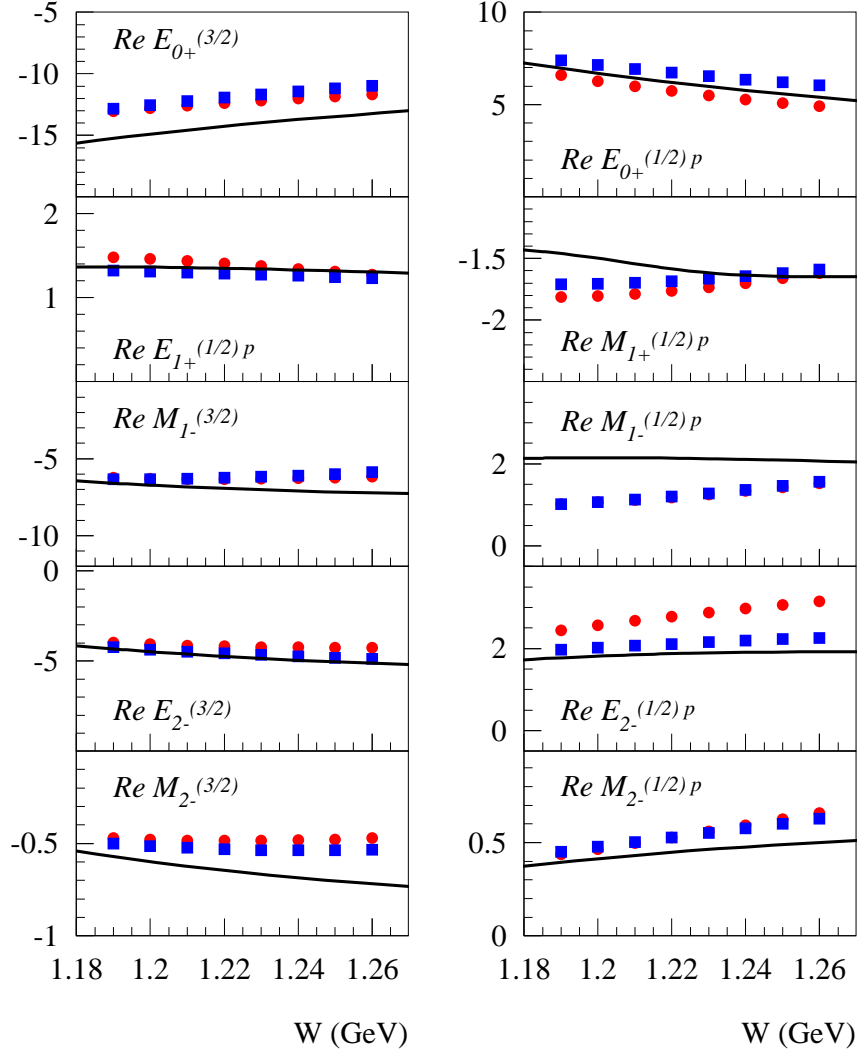}
\caption{
(Color online) Non-resonant multipole amplitudes
(in units $10^{-3} / m_\pi$) 
for pion photoproduction as function of the invariant mass $W$ of 
the $\pi N$ system.  
The solid curves result from our NLO calculation. 
The data points are from the 
SAID analysis (FA04K)~\protect\cite{Arndt:2002xv} (red circles), and from the 
MAID 2003 analysis~\protect\cite{Drechsel:1998hk} (blue squares).
}
\figlab{gap_pin_nonresmult}
\end{figure}

\begin{figure}[t,h]
\centerline{  \epsfxsize=14cm
  \epsffile{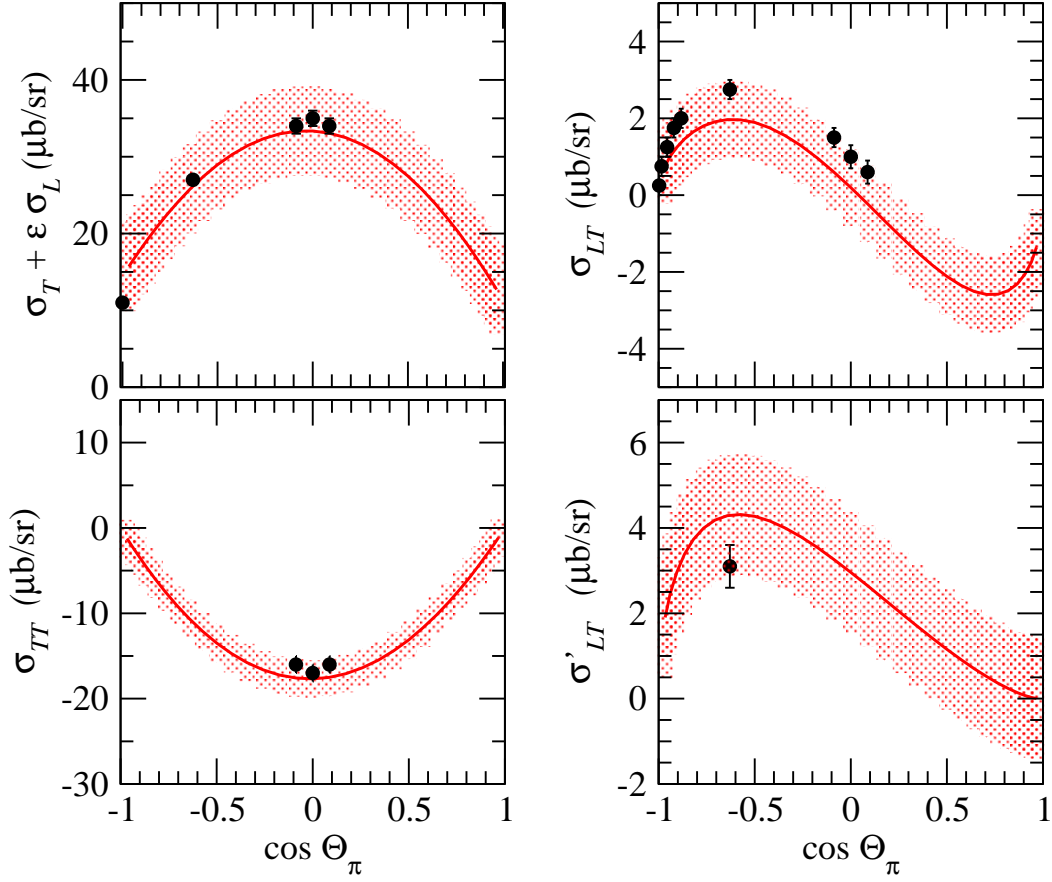} 
}
\caption{
(Color online) The NLO results for the  $\Th_\pi$ dependence 
of the $\ga^\ast p \to \pi^0 p$ cross sections at 
$W = 1.232$~GeV and $Q^2$ = 0.127~GeV$^2$. 
The theoretical error bands are  described in the text. 
Data points are from BATES experiments~\protect\cite{Bates01,Kunz:2003we}.
}
\figlab{crossections}
\end{figure}

\begin{figure}[t,h]
\includegraphics[width=0.7\columnwidth]{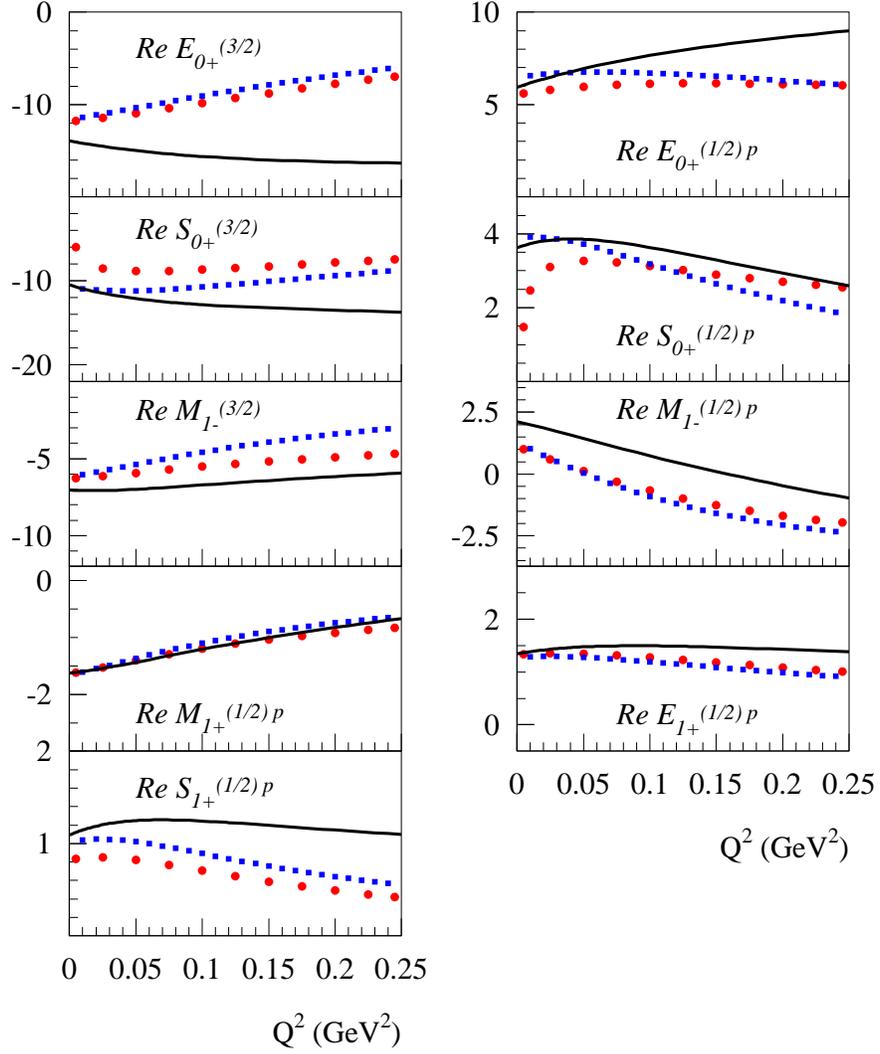}
\caption{
(Color online) Non-resonant multipole amplitudes
(in units $10^{-3} / m_\pi$) 
for pion electroproduction at the resonance position ($W = 1.232$~GeV) 
as function of $Q^2$.  
The curves are the results of our NLO calculation. 
The data points are from the 
SAID analysis (FA04K)~\protect\cite{Arndt:2002xv} (red circles), and from the 
MAID 2003 analysis~\protect\cite{Drechsel:1998hk} (blue squares).
}
\figlab{ep_epin_nonresmult}
\end{figure}

\begin{figure}
\centerline{  \epsfxsize=10cm%
\epsffile{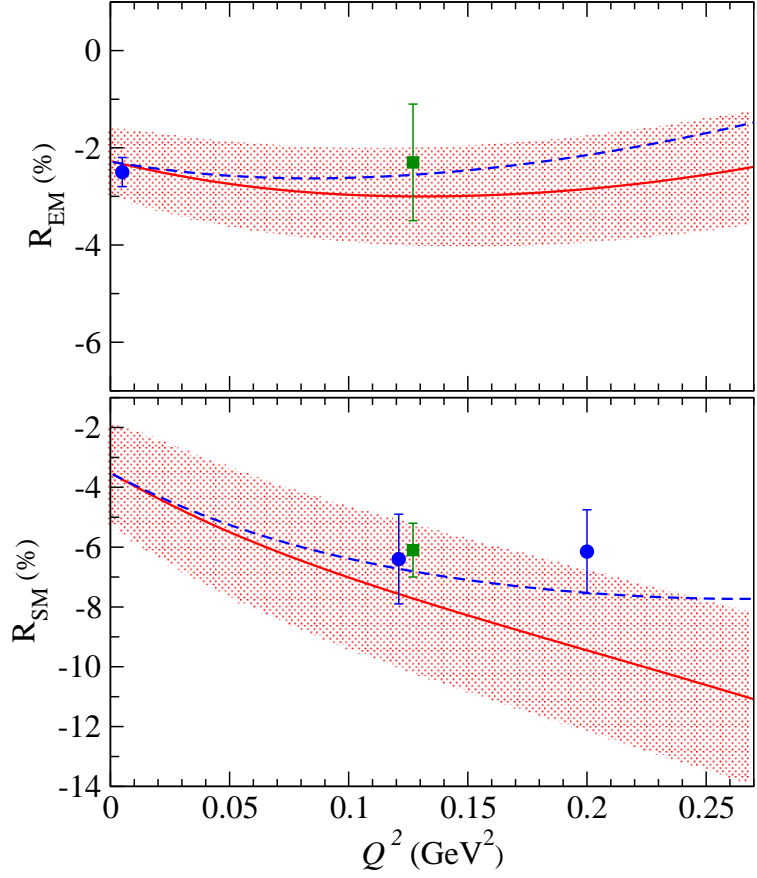} 
}
\caption{%
(Color online) $Q^2$ dependence of the NLO results (solid curves) 
for $R_{EM}$ (upper panel) and  $R_{SM}$ (lower panel). 
The blue 
dashed curves represent a phenomenological estimate of N$^2$LO effects by 
including $Q^2$-dependence in $g_E$ according to \Eqref{geqsqr}.   
The blue circles are data points from MAMI 
for $R_{EM}$~\protect\cite{Mainz97}, and 
 $R_{SM}$~\protect\cite{Pospischil:2000ad,Elsner:2005cz}.  
The green squares are data points from BATES~\protect\cite{Bates01}.
}
\figlab{ratiosQ2}
\end{figure}

\begin{figure}
\centerline{  \epsfxsize=12cm%
  \epsffile{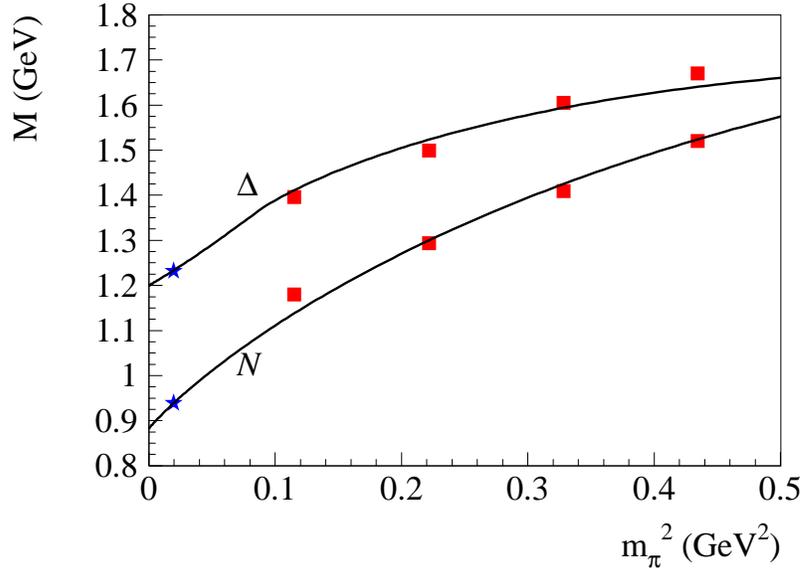} 
}
\caption{(Color online) Pion-mass dependence of the 
nucleon and $\Delta(1232)$ masses. 
The curves are two-parameter expressions for the $\pi N$ loop contributions 
to $M_N$ and $M_\Delta$ according to Eqs.~(\protect\ref{eq:nucpin}) and 
(\protect\ref{eq:delpin1}) respectively, using 
$M_N^{(0)} = 0.883$~GeV, $c_{1 N} = -0.87$~GeV$^{-1}$, 
and 
$M_\Delta^{(0)} = 1.20$~GeV, $c_{1 \Delta} = -0.40$~GeV$^{-1}$
respectively.  
The red squares are lattice results from the MILC 
Collaboration~\protect\cite{Bernard:2001av}. 
The stars represent the physical mass values. }
\figlab{nucdelmass}
\end{figure}


\begin{figure}[h,t]
\centerline{ \epsfxsize=10cm%
\epsffile{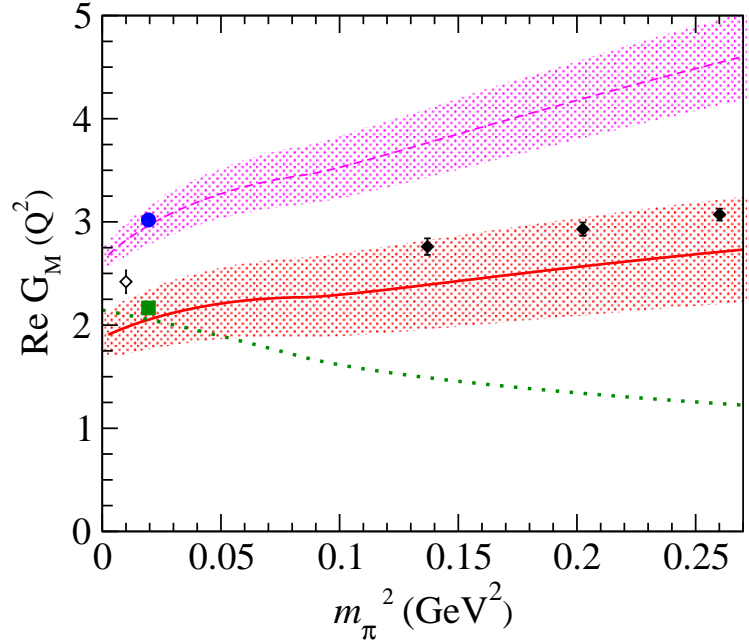}
}
\caption{%
(Color online) The $m_\pi$-dependence of the real part of 
the Jones-Scadron $\gamma N \Delta$ form factor $G_M^\ast$ for $Q^2 = 0$ and 
$Q^2$ = 0.127 GeV$^2$.
The solid (dashed) curves are the NLO results for $Q^2 = 0.127$~GeV$^2$
($Q^2 = 0$) respectively, including
the $m_\pi$ dependence of $M_N$ and $M_\Delta$. 
The green dotted curve is 
the corresponding result for $Q^2 = 0.127$~GeV$^2$ where   
the $m_\pi$ dependence of $M_N$ and $M_\Delta$ is not included. 
The blue circle for $Q^2 = 0$ 
is a data point from MAMI~\protect\cite{Mainz97} , and 
the green square for $Q^2 = 0.127$~GeV$^2$ is 
a data point from BATES~\protect\cite{Bates01}. 
The three filled black diamonds at larger $m_\pi$   
are lattice calculations~\protect\cite{Ale05} for $Q^2$ values of 
0.125, 0.137, and 0.144 GeV$^2$ respectively, 
whereas the open diamond near $m_\pi \simeq 0$ represents their  
extrapolation assuming linear dependence in $m_\pi^2$. 
}
\figlab{regmmpi}
\end{figure}

\begin{figure}
\centerline{  \epsfxsize=11cm%
  \epsffile{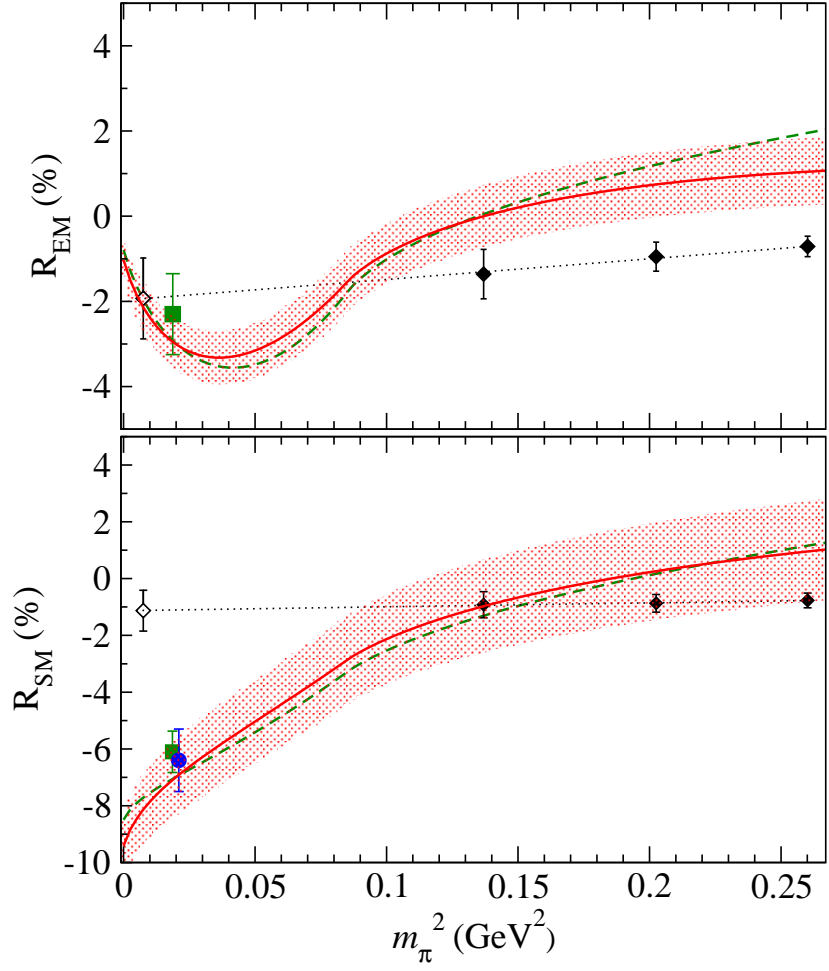} 
}
\caption{
(Color online) $m_\pi$ dependence of the NLO results at $Q^2=0.1$ GeV$^2$ for 
 $R_{EM}$ (upper panel) and 
$R_{SM}$ (lower panel).
The blue circle is a data point from MAMI~\protect\cite{Pospischil:2000ad}, 
the green squares are data points from BATES~\protect\cite{Bates01}. 
The three filled black diamonds at larger $m_\pi$   
are lattice calculations~\protect\cite{Ale05}, 
whereas the open diamond near $m_\pi \simeq 0$  
represents their extrapolation assuming linear dependence in $m_\pi^2$. 
Red solid curves: NLO result when accounting for the $m_\pi$ dependence in 
$M_N$ and $M_\Delta$; 
green dashed curves: NLO 
result of Ref.~\protect\cite{Pascalutsa:2005ts}, where   
the $m_\pi$-dependence of $M_N$ and $M_\Delta$ was not accounted for. 
}
\figlab{ratios}
\end{figure}

\end{document}